\documentclass[useAMS,usenatbib,fleqn]{mnras}

\usepackage{amsmath,amsfonts,amssymb}
\usepackage{booktabs}
\usepackage{graphicx}
\usepackage{pdflscape}
\usepackage{longtable}
\usepackage{deluxetable}
\usepackage[normalem]{ulem}
\usepackage{footnote}
\usepackage{multirow}

\newcommand{\OII}{[\mbox{O\,{\sc ii}}]}
\newcommand{\OIII}{[\mbox{O\,{\sc iii}}]}
\newcommand{\SiIV}{\hbox{{\rm Si}{\sc \,iv}}}
\newcommand{\NV}{\hbox{{\rm N}{\sc \,v}}}
\newcommand{\CIV}{\hbox{{\rm C}{\sc \,iv}}}
\newcommand{\HI}{\hbox{{\rm H}{\sc \,i}}}
\newcommand{\HII}{\hbox{{\rm H}{\sc \,ii}}}
\newcommand{\HeII}{\hbox{{\rm He}{\sc \,ii}}}

\title[Escape fraction of faint galaxies]{Constraints on the Lyman continuum escape fraction for faint star forming galaxies
\thanks{Based on observations collected at the European Southern Observatory for Astronomical research in the Southern Hemisphere
under ESO programme 094.A-0289(B).}}

\author[J. Japelj et al.]{
\parbox[t]{\textwidth}{J.~Japelj$^{1}$\thanks{E-mail: japelj@oats.inaf.it},
E.~Vanzella$^{2}$,
F.~Fontanot$^{1}$,
S.~Cristiani$^{1,3}$,
G.~B.~Caminha$^{4}$,
P.~Tozzi$^{5}$,
I. ~Balestra$^{6}$,
P.~Rosati$^{4}$,
M.~Meneghetti$^{2}$
}
\vspace*{8pt}\\
$^1$ INAF - Osservatorio Astronomico di Trieste, via G. B. Tiepolo 11, 34131 Trieste, Italy\\
$^2$ INAF - Osservatorio Astronomico di Bologna, via Ranzani 1, 40127 Bologna, Italy\\
$^3$ INFN - National Institute for Nuclear Physics, via Valerio 2, I-34127 Trieste, Italy\\
$^4$ Dipartimento di Fisica e Scienze della Terra, Universit\`a degli Studi di Ferrara, Via Saragat 1, I-44122 Ferrara, Italy\\
$^5$ INAF - Osservatorio Astrofisico di Arcetri, Largo E. Fermi, I-50125, Firenze, Italy\\
$^6$ University Observatory Munich, Scheinerstrasse 1, 81679 Munich, Germany
}

\date{Accepted XXX. Received YYY; in original form ZZZ}

\pubyear{2016}

\begin{document}
\label{firstpage}
\pagerange{\pageref{firstpage}--\pageref{lastpage}}
\maketitle

\begin{abstract}
Star forming galaxies have long been considered the dominant sources of the cosmic ultraviolet background radiation at early epochs. However, observing and characterizing the galaxy population with significant ionizing emission has proven to be challenging. In particular, the fraction of ionizing radiation that escapes the local environment to the intergalactic medium is poorly known. We investigate the relation between the escape fraction and galaxy luminosity. We combine deep ultraviolet observations of Hubble Ultra Deep Field (UVUDF) with deep Multi Unit Spectroscopic Explorer (MUSE) observations of the same field, collecting a sample of 165 faint star forming galaxies in the $3 < z < 4$ redshift range with deep rest-frame observations of the Lyman continuum. In our sample, we do not find any galaxy with significant emission of LyC radiation. We bin the galaxies in various redshift and brightness intervals and stack their images. From stacked images we estimate the relative escape fraction upper limits as a function of the luminosity. Thanks to the depth of the sample we measure meaningful 1$\sigma$ upper limits of $f_{esc,rel} < 0.07, 0.2$ and 0.6 at $L \sim L_{\rm z=3}^{*}, 0.5L_{\rm z=3}^{*}$ and $0.1L_{\rm z=3}^{*}$, respectively. We use our estimates and theoretical predictions from the literature to study a possible dependence of the escape fraction on galaxy luminosity by modelling the ionizing background with different prescriptions of $f_{\rm esc} (M_{\rm UV})$. We show that the understanding of the luminosity dependence hinges on the ability to constrain the escape fraction down to $M_{\rm UV} \sim -18$ mag in the future. \vspace{2cm}
\end{abstract}




\section{Introduction}
The epoch of cosmic reionization represents an important evolutionary stage of the Universe marking the formation of the first cosmic structures \citep[e.g.][]{Robertson2010}. A number of evidence suggests that reionization began at $z \sim 10 - 15$ and was completed by $z \sim 6$ \citep[e.g.][]{Robertson2015}. The details of the reionization process are however still poorly known. In particular, identifying the sources driving the ionization of neutral hydrogen during the reionization epoch remains a challenge. Our understanding of the physics of the intergalactic medium (IGM) and galaxy formation depends on the detailed knowledge of the cosmic background radiation, therefore it is important to characterize the nature of the ionizing sources.

According to the predominant view, star-forming galaxies are the main drivers of the reionization at high redshifts \citep[$z\gtrsim 4$; e.g.][]{Fontanot2012,Fontanot2014,Robertson2013,Robertson2015,Haardt2015,Cristiani2016}, although an identification of a numerous population of (faint) active galactic nuclei (AGNs) at $4 < z < 6.5$ \citep{Giallongo2015} indicates that the contribution of AGNs to the cosmic reionization at early epochs may be more important than previously assumed \citep{Madau2015,Khaire2016}. In order to understand the relative contribution of AGNs and star-forming galaxies to the cosmic ionization background through cosmic time we need to characterize the population of star-forming galaxies with considerable Lyman continuum (LyC) emission. Direct observations of galaxies are only viable up to redshifts $z\sim 3.5 - 4$, beyond which the rest-frame LyC emission becomes unobservable due to the increasing IGM opacity \citep[e.g.][]{Madau1995}. A viable strategy implies the search for $z < 4$ LyC-leaking galaxies and use their characteristic properties to identify high-redshift analogs, whose LyC cannot be directly observed \citep[e.g.][]{Zackrisson2013,Schaerer2016}.

The search for Lyman continuum leakers has been conducted by many surveys both at low \citep[$z < 1.5$;][]{Siana2007,Siana2010,Cowie2009,Grimes2009,Bridge2010} and high redshifts \citep[$z \sim 3 - 4$;][]{Steidel2001,Shapley2006,Iwata2009,Vanzella2010,Vanzella2012,Vanzella2015,Nestor2013,Mostardi2013,Guaita2016,Grazian2016,Marchi2016}. Only a handful of sources with significantly detected Lyman continuum emission have been identified among hundreds of inspected galaxies, after accounting for contamination of superimposed foreground sources, which becomes increasingly more important when moving towards higher redshifts and lower luminosities \citep{Vanzella2010,Mostardi2015,Siana2015}. Secure spectroscopically confirmed detections of the Lyman continuum from star-forming galaxies have been found in the local Universe \citep{Bergvall2006,Leitet2013,Borthakur2014}, at low redshifts \citep[$z\sim$0.3;][]{Izotov2016a,Izotov2016b} and high redshifts \citep[$z\sim3.2$;][]{Barros2016,Vanzella2016,Shapley2016}.  More photometric LyC candidates at $z \sim 2-3$ are awaiting confirmation \citep[e.g.][]{Mostardi2015,Naidu2016}.

In order to assess the contribution of star-forming galaxies to the cosmic ionization background we need to know their number densities, their ionizing photon production efficiency (i.e. the number of LyC photons per UV luminosity; \citealt{Bouwens2016}) and the fraction of the LyC photons (i.e. escape fraction $f_{\rm esc}$) that can actually escape the local environment and ionize the IGM \citep{Robertson2013}. The escape fraction is the least constrained among the three quantities. Values for the galaxies with detected LyC (see above) range from a few \citep{Leitet2013} to $>50\%$ \citep{Vanzella2016}. Because the frequency of detection is low, the escape fraction is low on average: analysis of large galaxy samples shows that on average $f_{\rm esc} < 0.05$ at $3 < z < 3.5$ \citep{Vanzella2010,Grazian2016}. Assuming that star-forming galaxies are the only contributors to the ionization at $z\gtrsim 5$, the globally averaged escape fraction required to keep the Universe ionized should be $\sim 20\%$ \citep{Ouchi2009,Bouwens2012,Finkelstein2012,haardt2012,Khaire2016}. This value is obtained by extrapolating the UV luminosity function to $M_{\rm UV} \sim -17$ mag, while the galaxy samples being used to search for Lyman continuum are limited to relatively bright galaxies ($L\sim L_{\rm z=3}^{*}$; \citealt{Vanzella2010,Mostardi2015}). The galaxies with significantly detected Lyman continuum are in the same brightness range \citep{Vanzella2016,Schaerer2016}. It has been therefore suggested that the escape fraction (on average) may evolve with redshift and/or luminosity \citep[e.g.][]{Ciardi2012,Kuhlen2012,Fontanot2014}. Together with the contribution from AGNs, such prescription is successful in reproducing the observed properties of the background in the $3 < z < 5$ range \citep[e.g.][]{Becker2013}. 

In this work we combine the {\it HST} WFC3-UVIS/F336W deep ultraviolet observations of the Hubble Ultra Deep Field \citep[UDF;][]{Beckwith2006} with {\it VLT}/MUSE deep observations of the same field, collecting a sample of $\sim 165$ faint galaxies in the $3 < z < 4$ range with deep rest frame $<912\mathrm{\AA}$ observations. Our aim is to search for galaxies with possible Lyman continuum emission and to provide upper limits on the $f_{\rm esc}$ in the $L > 0.1L_{\rm z=3}^{*}$ luminosity range. The presentation of data and the selection of the sample are given in Section \ref{selection}. Results are reported in Section \ref{results}. In Section \ref{discuss} we investigate the possibility of a luminosity-dependent escape fraction by modelling the ionizing background using different $f_{\rm esc}(M_{\rm UV})$ prescriptions. All magnitudes are quoted in the AB photometric system. We use standard cosmology \citep{Planck2014}.

\section{Data and sample selection}
\label{selection}

\subsection{Ultra-violet Hubble Ultra Deep Field}
\label{uvudf}

\begin{figure}
\centering
\includegraphics[scale=0.55]{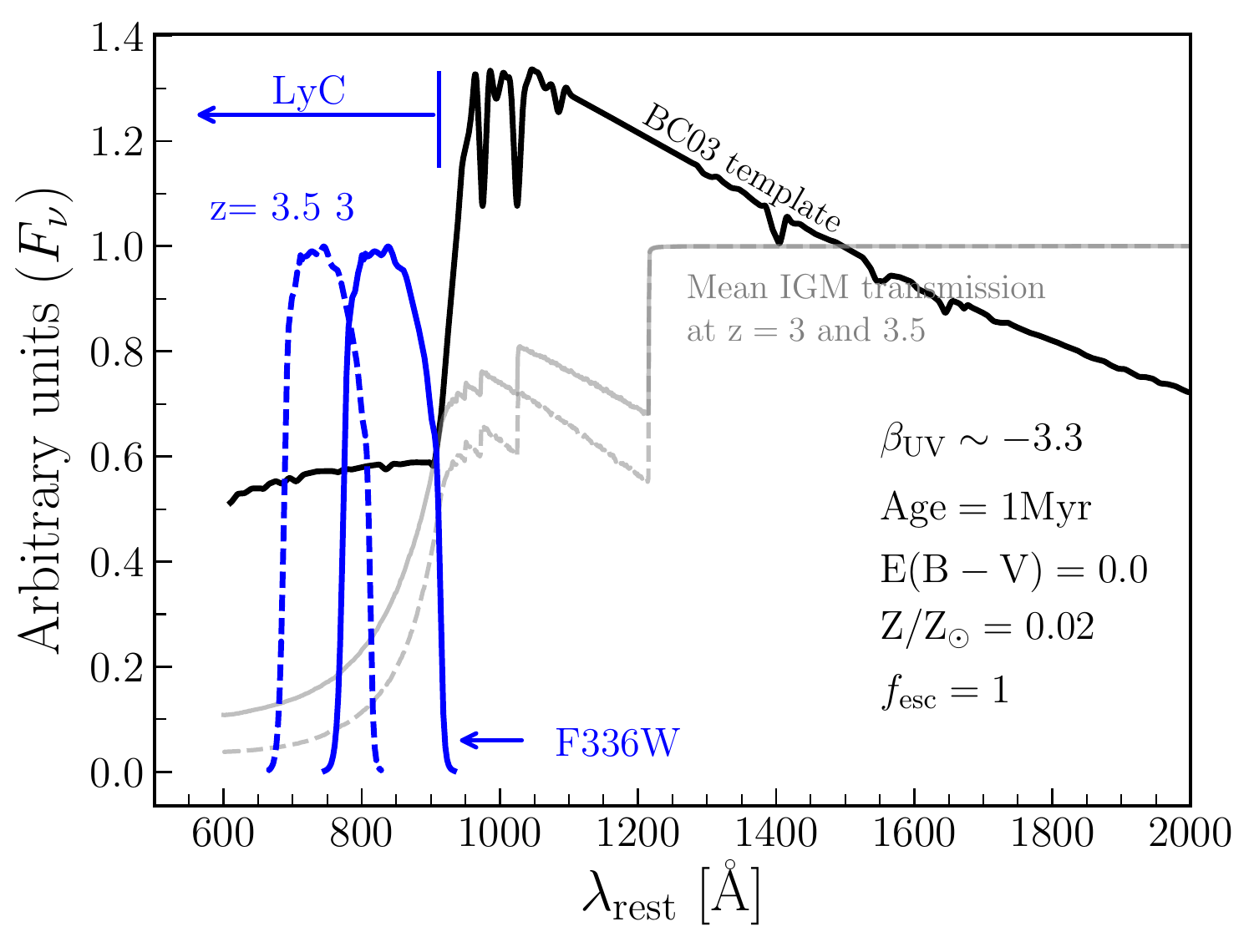}
\caption{Example of a galaxy template spectrum in the system of the galaxy (black line). Overplotted are the normalized F336W filter transmission curves assuming we observe a galaxy at redshift $z = 3$ (solid blue) or $z = 3.5$ (dashed blue). Mean IGM transmission at $z = 3$ (solid grey) or $z = 3.5$ (dashed grey) is also shown.}
\label{fig0}
\end{figure}

To investigate the Lyman continuum emission of galaxies at $z \sim 3 - 4$, we need deep imaging with filters probing the rest-frame emission bluewards of the Lyman limit. We use observations of a subfield of the UDF, which has been recently acquired in the ultra-violet (UVUDF; \citealt{Teplitz2013,Rafelski2015}) with three WFC3-UVIS filters: F225W, F275W and F336W\footnote{\url{http://uvudf.ipac.caltech.edu/}}. For our purposes we use the observations obtained with the F336W filter, because it probes the $\lambda_{\rm rest} < 912 ~\mathrm{\AA}$ spectral region for galaxies at $z\gtrsim 3$. We show this schematically in Fig. \ref{fig0}, where a synthetic spectrum of a young, low-metallicity galaxy with $f_{\rm esc} = 1$ (see \citealt{Vanzella2015} for details) is plotted together with the normalized F336W filter transmission curve. In addition to the deep ultra-violet observations, the advantage of using this field is the availability of the imaging at longer wavelengths and high spatial resolution \citep{Beckwith2006}, which is crucial to identify galaxies that are free of contamination from interlopers \citep{Vanzella2010,Siana2015,Mostardi2015}.

\begin{figure}
\centering
\includegraphics[scale=0.45]{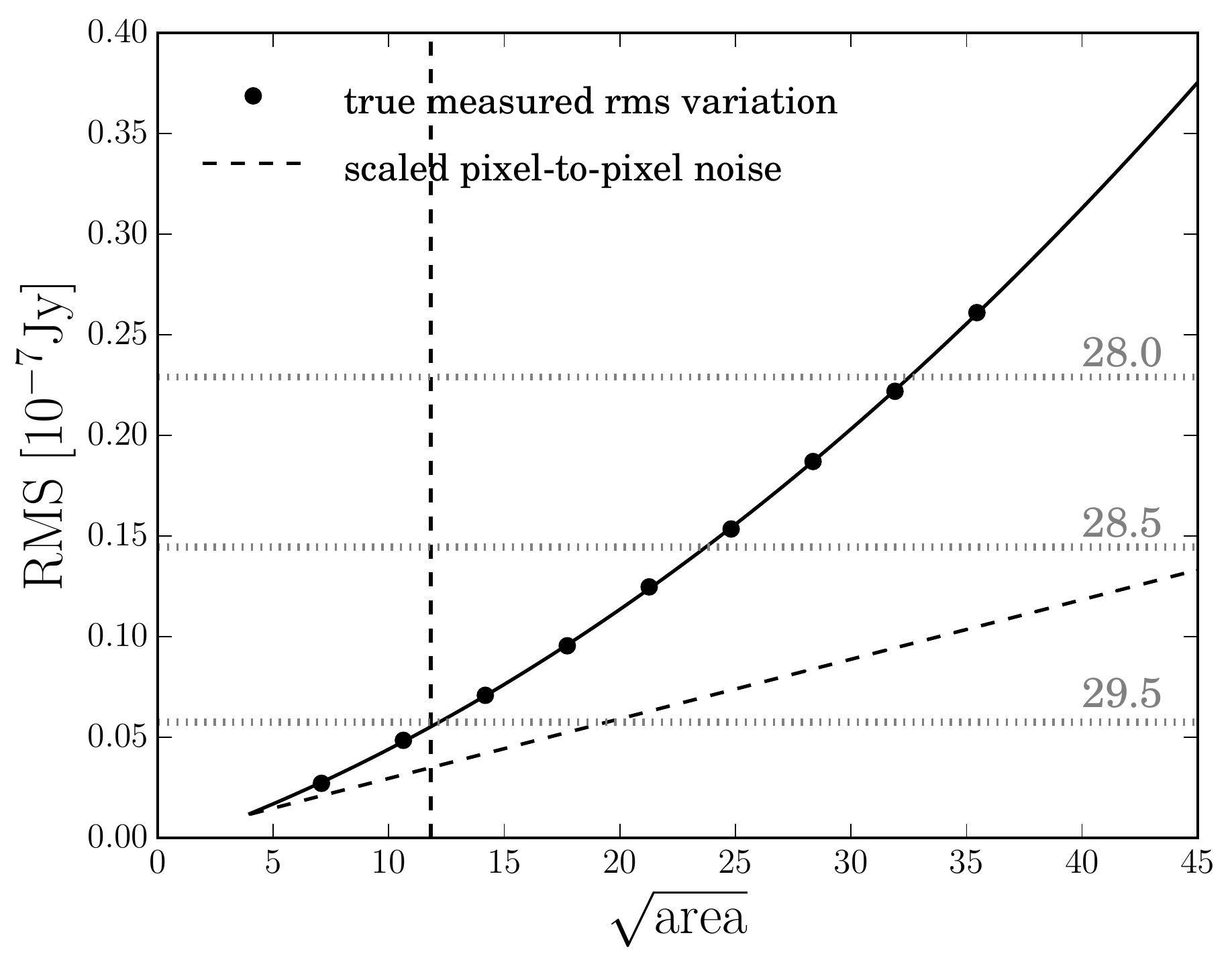}
\caption{Measurement of the depth of the (30 mas/pixel) drizzled image of the UVUDF field taken in UVIS/F336W filter \citep{Rafelski2015}. RMS is shown as a function of $\sqrt{A}$ of the aperture with area $A$ (in pixels). Horizontal dotted lines indicate the corresponding 1$\sigma$ upper limits (in magnitudes). Dashed line: assuming uncorrelated noise, pixel-to-pixel variation $(\sigma_{i})$ is scaled to the desired aperture as RMS$ = \sigma_{i}\sqrt{A}$. This results in a $1\sigma$ $(5\sigma)$ limiting magnitude in the 0\farcs2 aperture of 30.0 (28.3) mag. Black circles: directly measured flux variations in circular apertures of different sizes. Full line is a polynomial fitted to the points and does not represent a physical model. This method results in a $1\sigma$ $(5\sigma)$ limiting magnitude of 29.5 (27.8) mag. Vertical dashed line corresponds to 0\farcs2 aperture. Measurements are obtained using Monte Carlo simulations and only in regions of the image where no sources have been detected.}
\label{fig1}
\end{figure}

An important quantity used throughout the study is the depth of the F336W image. \citet{Rafelski2015} quote $5\sigma$ upper limit of 28.3 mag for an aperture of 0\farcs2. This value was obtained by measuring a pixel-to-pixel noise and scaling it to the 0\farcs2 aperture, assuming the noise is uncorrelated. While we reproduce this value following the steps of \citet{Rafelski2015}, we also measure the flux variations in the image background -- where no sources are detected -- directly in apertures of different sizes. Flux within each aperture is measured at 1000 random positions in the image. RMS of the image is estimated as sigma-clipped standard deviation of the resulting distribution of fluxes. In this way, we measure a slightly lower $5\sigma$ upper limit of 27.8 mag. The measurement is illustrated in Fig. \ref{fig1}. In the following we adopt the latter as a more conservative value.

\begin{figure}
\centering
\includegraphics[scale=0.55]{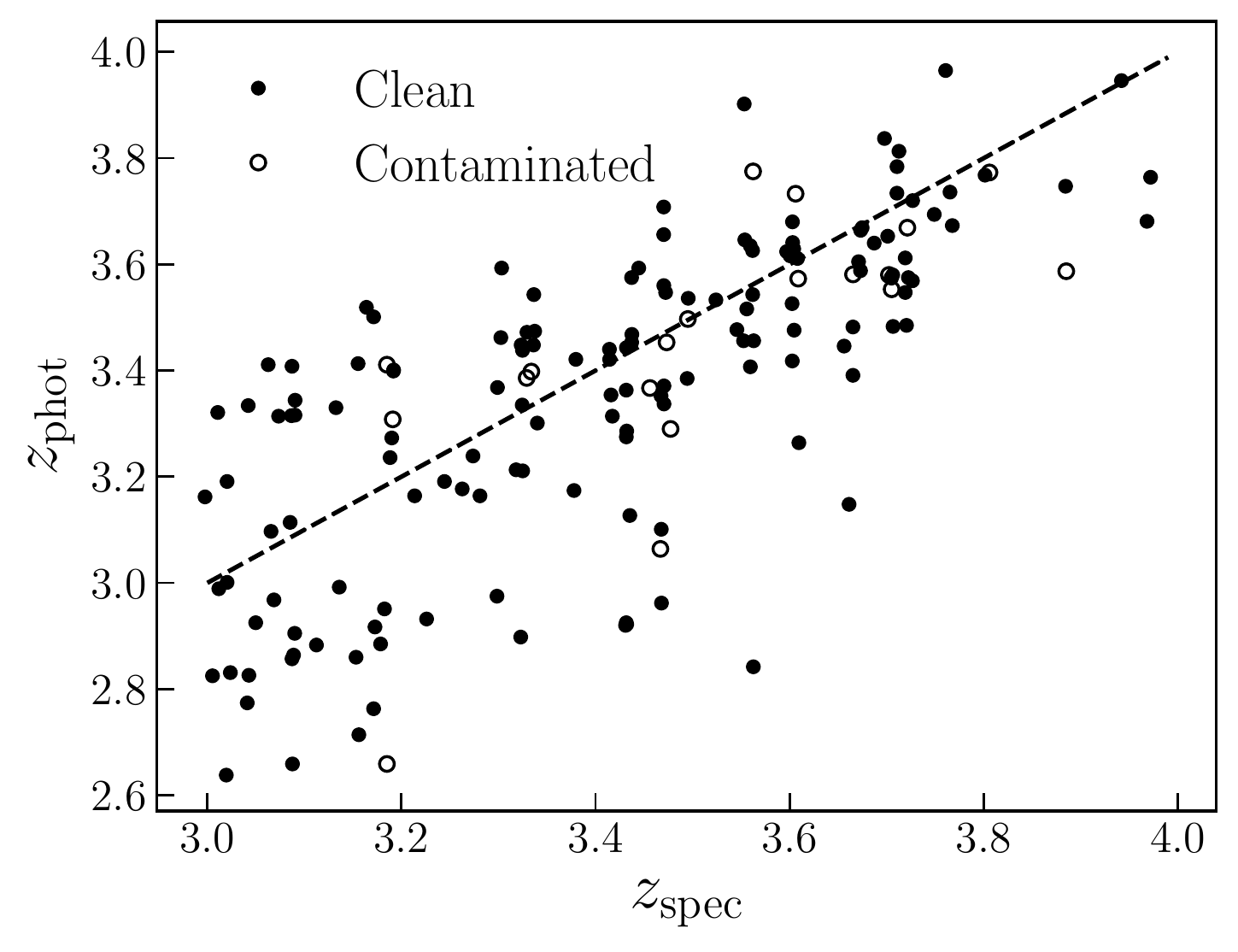}
\caption{Comparison between photometric redshifts \citep{Coe2006} and Ly$\alpha$-based spectroscopic redshifts for a sample of 165 galaxies identified in the $3 < z < 4$ range in the UVUDF field. Galaxies referred to as contaminated are suspected to have interlopers in the line of sight. The dashed line shows $z_{\rm spec} = z_{\rm phot}$ relation.}
\label{fig1a}
\end{figure}

\subsection{MUSE deep observations}

The Multi Unit Spectroscopic Explorer \citep[MUSE][]{Bacon2010} is a second-generation VLT panoramic integral-field spectrograph. It has a field of view of $1 \time 1$ arcmin$^2$, spatial resolution of 0\farcs2, spectral range of 4750-9350 $\rm \AA$ and resolution of R $\approx 3000$, making it a very efficient instrument for simultaneous observation of a large number of faint galaxies. We used MUSE observations of the UDF field conducted between September and December 2015 under the  GTO Program ID. 094.A-0289(B) (PI: R. Bacon). The data consist in a mosaic covering $3 \times 3$ arcmin$^2$ of 9 pointings, plus one deeper pointing (UDF-10) in the central region of the field (overlapping the mosaic). We have used the archival data that are publicly available to date, i.e. $\approx 6$ hours of exposure time for each pointing in the $3 \times 3$ mosaic, and the full 20 hours in the UDF-10 field of view. Each exposure is 25 minutes, totalling 180 pointings, rotated by 90$^o$, with small offsets of a fraction of arcsecond to reduce observational systematics, such as residuals in the sky subtraction and instrumental effects. The overall observational conditions were good with a mean DIMM seeing of $0\farcs85 \pm 0.28$.

We used the MUSE reduction pipeline version 1.2.1 \citep{Weilbacher2012,Weilbacher2014} to process the raw data and create the final data-cubes. All the standard calibrations provided by the pipeline were applied in each exposure (bias and flat field corrections, wavelength and flux calibration, etc.), following the same procedure detailed in our previous works \citep{Caminha2016,Vanzella2016b}. Subsequently, we combined the calibrated observations into the 10 data-cubes covering the UDF area. Each data-cube has a spatial pixel scale of $0\farcs2$ and spectral coverage from 4750 to 9350 ${\rm \AA}$, with a spectral scale of $1.25 \rm \AA/pixel$ and a fairly constant spectral resolution of $\approx 2.4 \rm \AA$. The final coordinate match were done matching the source positions detected with SExtractor \citep{Bertin1996} in the collapsed MUSE 2D images with the corresponding deep F606W HST observations \citep{Illingworth2013}, resulting in a residual rms smaller than the $0\farcs2$. Finally we applied the Zurich Atmosphere Purge \citep[ZAP;][]{Soto2016} to reduce the remaining sky residuals, using SExtractor segmentation maps to mask the objects with detected emission in the field-of-view. The final dataset consists of 10 MUSE cubes with some spatial overlaps covering an area of $3 \times 3$ arcmin$^2$. We checked the spectra of sources with detection in more than one data-cube, finding excellent consistence of wavelength and flux calibration.

To search for the suitable galaxy candidates for our work we employ the following technique. As a parent sample we take the catalogue of \citet{Coe2006} who provide photometric redshifts of the objects in the UDF field. We select all the objects within the UVUDF field with $2.5 < z_{\rm phot} < 4.0$, resulting in $\sim 2000$ candidate galaxies. Then we search for these objects in the MUSE data cubes and investigate their spectra in order to measure their redshifts. Our work is focused towards faint galaxies: with the used instrumental setup and the integration times, the spectra of $R \gtrsim 25$ mag do not have sufficient signal to noise to allow us to measure absorption-line-based redshifts. Instead we have to rely on detections of Ly$\alpha$ emission lines which means that our sample is largely selected of Ly$\alpha$ emitters: there are only two galaxies in the bright part of the sample with no significant Lya emission for which we were able to securely measure their redshifts from absorption lines. Furthermore, the fainter the object we are observing, the more we are biased towards systems with strong Ly$\alpha$ emission. Given the current depth of the survey we estimate that only lines with the rest-frame EW(Ly$\alpha$) $> 60 \mathrm{\AA}$ can be detected with 3-sigma significance for objects fainter than $\sim28$ mag. We emphasize that, while in the following we report all the measurements in the $3 < z_{\rm spec} < 4$ range (in the following we use $z_{\rm spec} = z$), we consider only objects brighter than $28.5$ mag in the stacking analysis of Section \ref{stack}. We measure the redshifts by fitting a simple Gaussian function to Ly$\alpha$ emission lines. In cases where Ly$\alpha$ emission is double-peaked, we fit a gaussian to each peak and measure the average redshift. We caution that, on average, Ly$\alpha$ lines are known to have offsets of several hundred km s$^{-1}$ with respect to systemic redshifts of galaxies \citep[e.g.][]{Pettini2001}. We also do not attempt to make a detailed classification of otherwise complex Ly$\alpha$ morphology (see e.g. \citealt{Kulas2012}) in this work. A more thorough account of the redshifts of all the objects in the field will be given in a subsequent work.

Thanks to the great efficiency of MUSE in detecting emission lines and and unambiguously distinguish their nature (for example OII$\lambda3726,3729$, CIV$\lambda1548,1551$, etc., see \citealt{Caminha2016} and \citet{Drake2016}, we securely identify Ly-$\alpha$ emission in 213 galaxies in the UVUDF field of view. Limiting ourselves to the $3.0 < z < 4.0$ range, the number drops to 165 galaxies. In Fig. \ref{fig1a} we show the relation between the spectroscopic redshift (measured in this work) and photometric redshift from \citet{Coe2006}. We have a reason to believe that a small part of the sample is photometrically contaminated, i.e. that there is another galaxy lying close to the line of sight of the galaxy for which we measure the spectroscopic redshift (for details see the next Section). From Fig. \ref{fig1a} it is apparent that there is no systematic offset for the contaminated galaxies. Finally, in the considered redshift range, \citet{Coe2006} did not have observations of rest-frame Lyman continuum emission. Their redshift measurements are therefore not affected by the strength of the LyC emission. \citet{Vanzella2015} also showed that, even if the escaping LyC from high-$z$ galaxies is taken into account, the photometric selection is not compromised.

Our sample is largely made of Lya emitters. It is currently unclear how strongly the escape fraction of the LyC depends on the strength of the Ly$\alpha$ line, though simulations have shown that escape fractions of Ly$\alpha$ and LyC radiation are expected to be positively correlated \citep{Behrens2014,Dijkstra2016}. The majority of the galaxies with the detected LyC also have strong Ly$\alpha$ emission\footnote{A strong LyC candidate awaiting for a secure redshift measurement, the {\it Ion1} galaxy found by \citet{Vanzella2012}, has no Ly$\alpha$ emission detected.} \citep{Verhamme2016}. However in previous studies hundreds of LBG galaxies, with all types of Ly$\alpha$ emission, have been inspected for LyC resulting in low upper limits for the escape fraction. Since galaxies with no Ly$\alpha$ emission are more numerous in such samples, our results should not be strongly affected by our selection. The results of this paper should be understood with this caveat in mind.

\begin{figure*}
\centering
\includegraphics[scale=0.35]{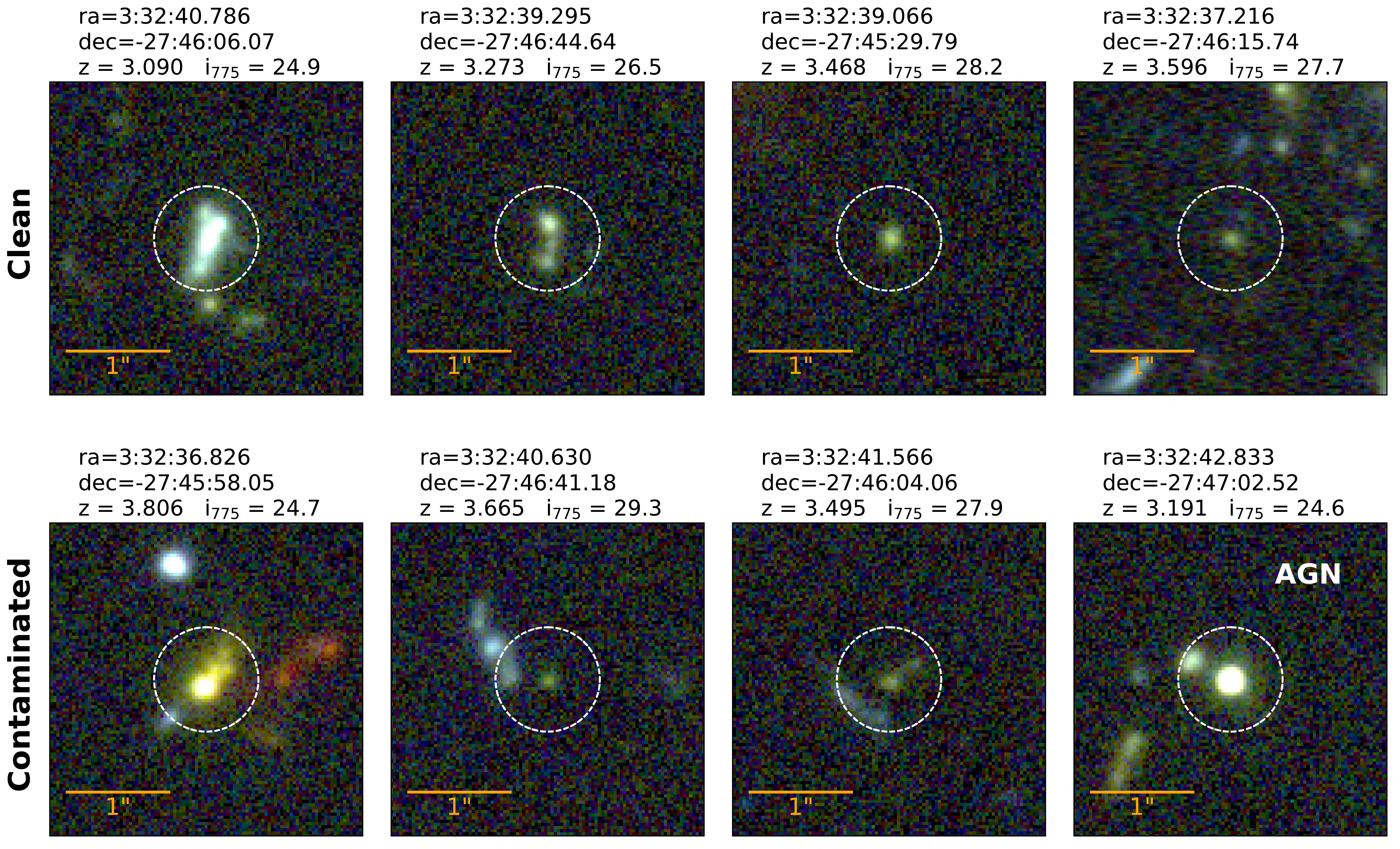}
\caption{Representative galaxies in the {\it clean} (top) and {\it contaminated} (bottom) samples. $3\arcsec\times3\arcsec$ colour images are stacked BVi images obtained with the {\it HST} ACS camera. Dashed circle is 0\farcs5 radius aperture. One of the galaxies has been identified as an AGN. For each galaxy we report the position (i.e. the centre of the white aperture), measured redshift and $i_{755}$ magnitude. The latter is taken from the catalogue of \citet{Rafelski2015}.}
\label{fig2}
\end{figure*}

\subsection{Clean galaxy sample}
\label{class}

All of these 165 galaxies have available {\it HST}/ACS observations, enabling us to visually inspect individual galaxies for any contaminants. We conservatively decided to work only with galaxies whose distance to the closest neighbour is more than 0\farcs5. We use the {\it HST}/ACS colour images of the field and inspect the region around each galaxy in the sample. If the source is clumpy and the colours of all the clumps are indistinguishable, we consider the source as clean from contamination. This decision is further corroborated by resolved emission-line region in the MUSE images extending over all the clumps. On the other hand, different colours are likely due to sources lying at different redshifts and are therefore assumed to be contaminated. We also checked for possible AGN emission. We do this both by cross-matching the positions of our sources with the catalogue from the Chandra Deep Field Survey \citep[CDFS;][]{Luo2008,Luo2016} and by looking for AGN spectral signatures in the MUSE spectra (e.g. \NV, \SiIV\, and \CIV\, emission lines). Only one source among our galaxies is found in the CDFS catalogue. Overall, 20 galaxies in the $3.0 < z < 4.0$ range are identified as being potentially contaminated. The remaining {\it clean} sample of 145 galaxies is used for further analysis. Images of a few representative galaxies of the {\it clean} and {\it contaminated} samples are shown in Fig. \ref{fig2}.

\begin{figure}
\centering
\includegraphics[scale=0.42]{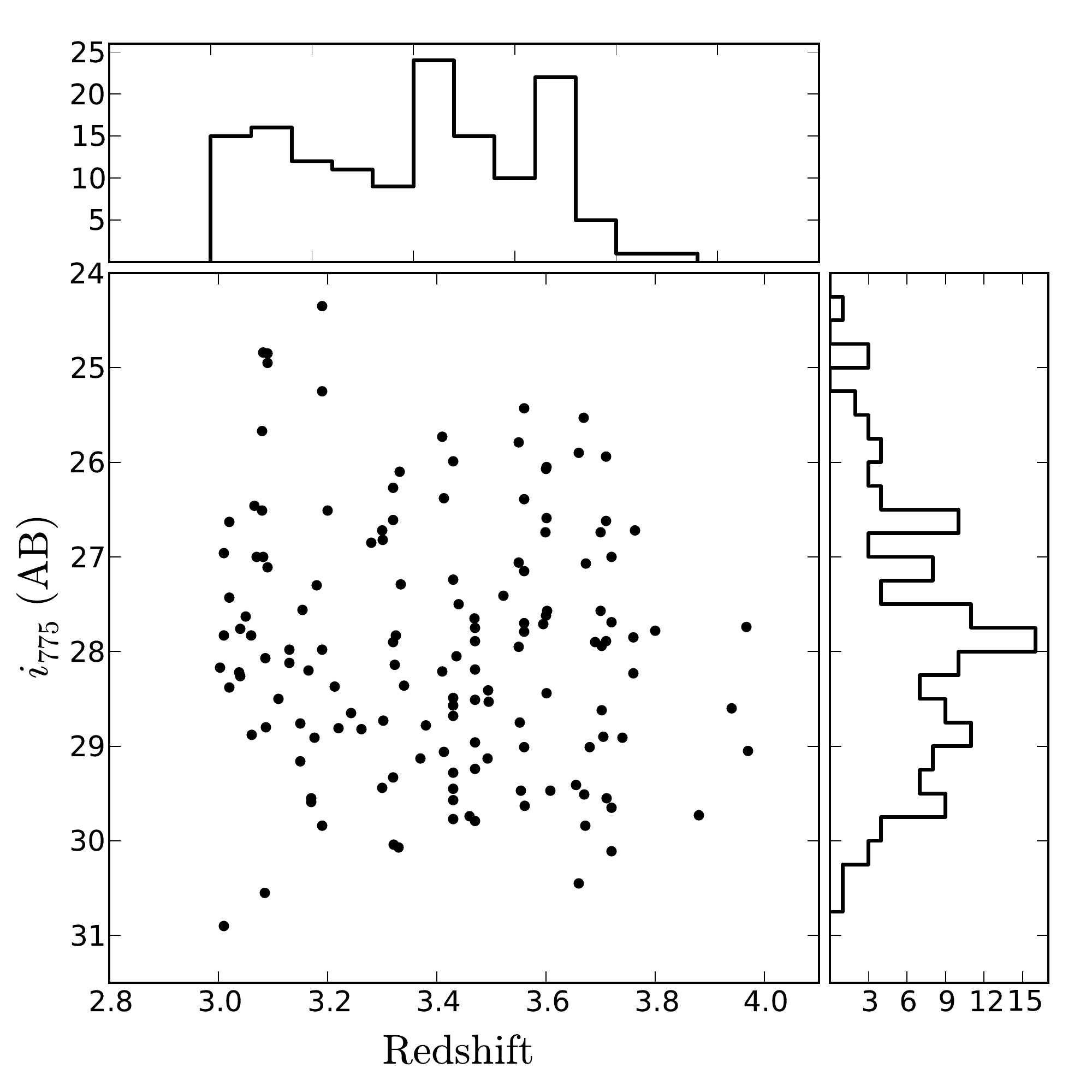}
\caption{2D and 1D distributions of the spectroscopic redshifts and HST/ACS F775W magnitudes of the 145 galaxies in the {\it clean} sample.}
\label{fig3}
\end{figure}

We search the literature for additional information on the galaxies of the {\it clean} sample. In particular, \citet{Rafelski2015} report photometry obtained with several {\it HST} cameras in the ultra-violet to near infrared (i.e. $\sim 0.2 - 1.6~\mathrm{\mu m}$) spectral range. We complement this data set with photometry given by \citet{Guo2013}, who extended the measurements to the near infrared where available. The catalogue of \citet{Guo2013} does not include sources as faint as the one of \citet{Rafelski2015}, therefore the near infrared measurements are not available for the whole {\it clean} sample. For the bright part of the sample we also retrieve measurements of stellar masses from \citet{Santini2015}. General properties of the galaxies in the sample are presented in Fig. \ref{fig3}. The sample is homogeneous in terms of redshift and brightness in the $3.0 < z < 3.8$ interval. We note that the galaxies are fainter with $\Delta \mathrm{m} \simeq 3$ mag on average with respect to the sample of \citet{Vanzella2010} in the same redshift range.

\section{Methodology and results}
\label{results}

Having defined the sample, we first measure fluxes in the F336W image at the positions of the galaxies. For compact galaxies we measure the flux within the 0\farcs2 aperture, as in Section \ref{uvudf}. However, many galaxies show a complicated substructure, a non-compact morphology, and are treated with the following approach. The stacked {\it HST}/ACS image is used to define an aperture encompassing all the pixels whose flux value is above 1.5 times the local RMS of the background around the galaxy. The F336W flux of the non-compact galaxy is then measured adopting this aperture. This approach should suffice for our proposes in this work since we are only interested whether the flux at the galaxy position is significant/detectable, however we caution that this methodology might not be desirable/good if precise photometry is necessary.

\begin{figure}
\centering
\includegraphics[scale=0.57]{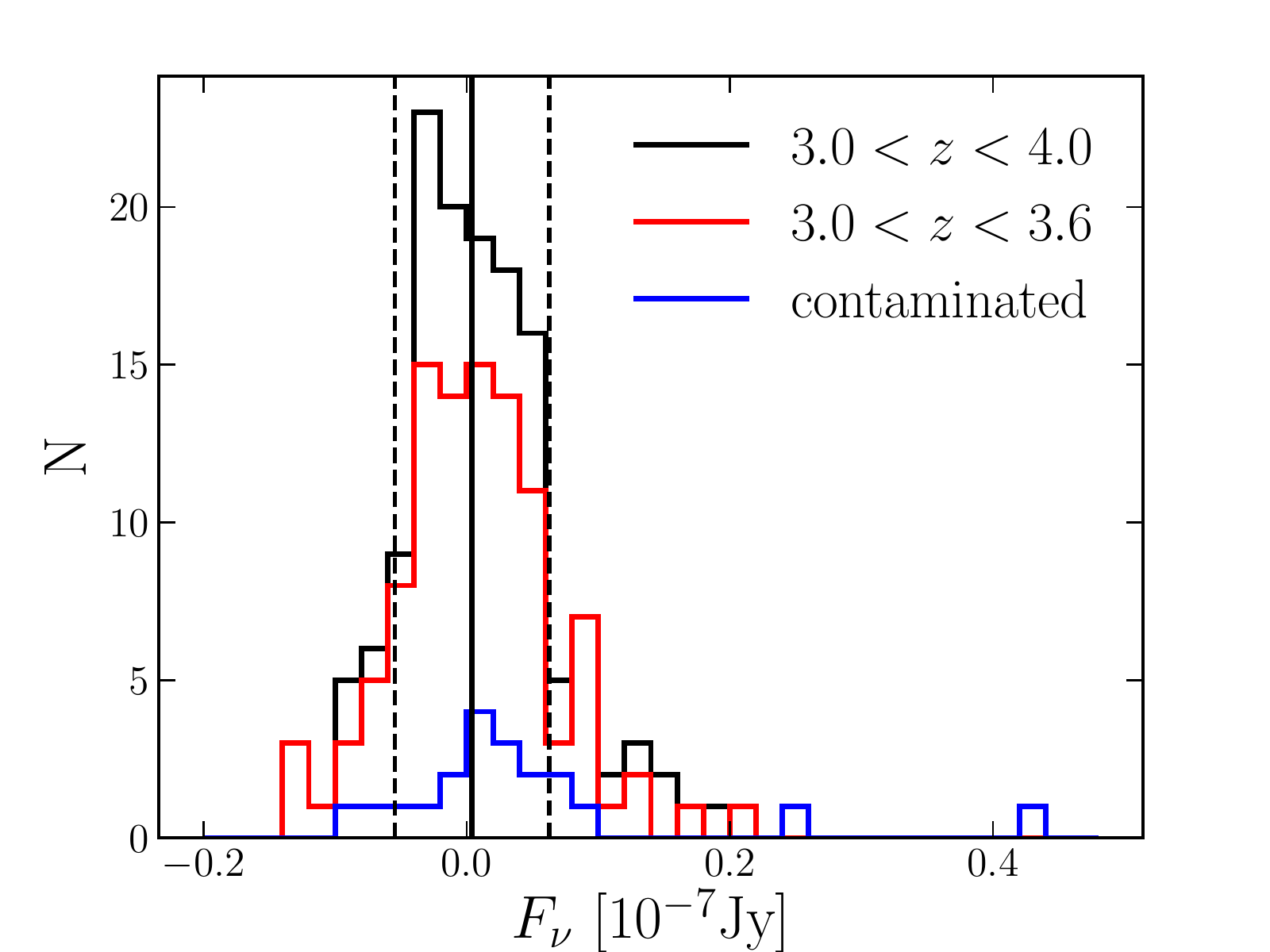}
\caption{UVIS/F336W fluxes measured at the galaxy positions (see text for details). Mean value and standard deviation are shown with full and dashed line, respectively. These values are very similar for both $3 < z < 4$ as well as $3 < z < 3.6$ range. Emission may have potentially been detected with $> 3\sigma$ significance in 4 cases - see Fig. \ref{fig5} for more details on their properties.}
\label{fig4}
\end{figure}

The distribution of fluxes measured in the described way is shown in Fig. \ref{fig4}. The fluxes of the {\it clean} sample have sigma-clipped mean and standard deviation of $0.004 \pm 0.056 \times 10^{-7}$ Jy. This $1\sigma$ dispersion corresponds to the magnitude of 29.5 mag, which is consistent with the value derived in Section \ref{uvudf} (see also Fig. \ref{fig1}). As the mean IGM transmission rapidly decreases with redshift (see discussion below), we also check the distribution of fluxes from galaxies limited to the $3.0 < z < 3.6$ range. We find that the mean and standard deviation do not change. For two galaxies ({\it G1} and {\it G2}) we find a potential detection with $\sim 3\sigma$ significance. Furthermore, we repeat the measurements for the 20 galaxies from the {\it contaminated} sample. Fluxes for this sample are similarly crowded around zero, but there are two notable detections ({\it GC1} and {\it GC2}) with high ($\sim 4$ and $7.5\sigma$) significance.

\begin{figure*}
\centering
\begin{tabular}{c}
\includegraphics[scale=0.32]{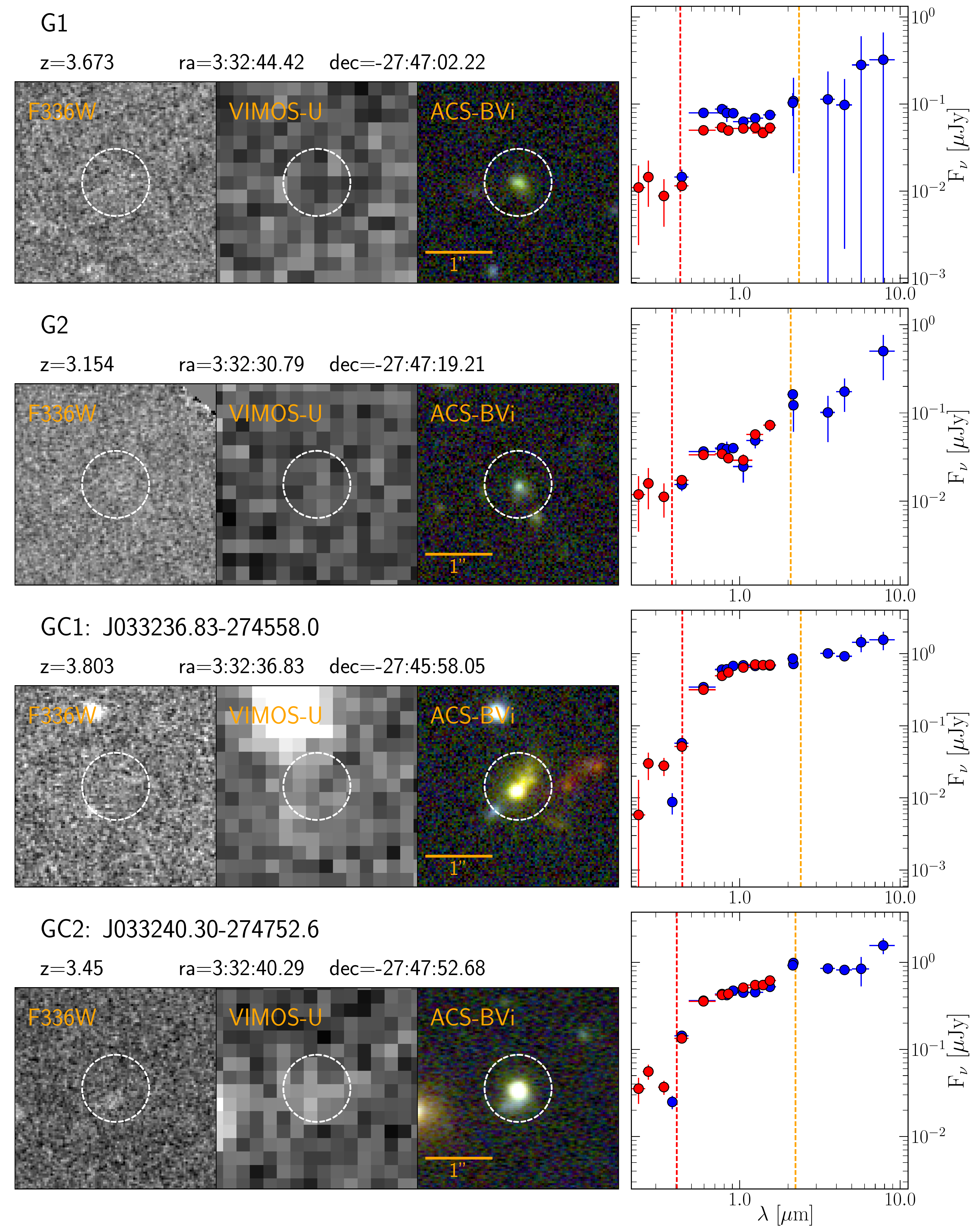}
\end{tabular}
\caption{Detailed look into the observational properties of the four galaxies with detected signal in F336W filter at $\gtrsim 3\sigma$ significance. Top two galaxies are from our {\it clean} sample, while the bottom two are from {\it contaminated} sample. $3\arcsec\times3\arcsec$ images show observations in {\it HST} WFC3/F336W and ACS colour images. We also include the deep ground based {\it VLT} VIMOS-U images \citep{Nonino2009}. White dashed circles show 0\farcs5 radius aperture. Plotted are also spectral energy distributions, using photometric measurements from \citet{Rafelski2015} (red points) and \citet{Guo2013} (blue points). Vertical red and orange dashed lines indicate rest-frame wavelengths of 912 and 5007 $\mathrm{\AA}$, respectively. We also quote the central positions and, where available, GOODS designations \citep{Giavalisco2004}.}
\label{fig5}
\end{figure*}

In Fig. \ref{fig5} we show the images in different wavelengths of the four galaxies with flux detected at $\gtrsim 3\sigma$. We also include their spectral energy distributions (SEDs) built over a wide spectral range. Galaxies {\it G1} and {\it G2} from the {\it clean} sample are moderately faint ($i_{775}\sim 27.3$ mag), while {\it GC1} and {\it GC2} from the {\it contaminated} sample are relatively bright ($i_{775}\sim 24.7$ mag). Stellar masses of $\log M_{\star}/M_{\odot} \sim 8.6, 9.2, 9.4$ and 9.7 are obtained from the SED modelling \citep{Santini2015} for {\it G1}, {\it G2}, {\it GC1} and {\it GC2}, respectively. Galaxies {\it G2} and {\it GC2} show an excess in the {\it K}-band flux, suggesting that strong $H\beta + \OIII\lambda4959,5007$ emission lines enter the {\it K}-band at their respective redshifts. \OIII\ lines are redshifted redwards of the {\it K}-band in the case of other two galaxies. A large {\it K}-band flux excess with respect to the flux measured in the {\it H}-band implies a possibly large \OIII/\OII\ ratio. The observed excess is interesting in view of the proposed connection between large \OIII/\OII\ and high escape fraction of Lyman continuum \citep[e.g.][]{Jaskot2013,Nakajima2014,Nakajima2016}. We discuss this more thoroughly in Section \ref{excess}.

All four galaxies were also detected in the F336W image by \citet{Rafelski2015} using proper aperture-matched PSF corrected photometry, though we note that detections are of low significance in all cases with the exception of {\it GC2}. Furthermore, the galaxies {\it G1} and {\it G2} are not detected in the VIMOS {\it U} image. This image is deeper than the one obtained with the F336W and the {\it U} filter covers slightly redder wavelengths (e.g. see Fig. 1 in \citealt{Vanzella2010}). The non-detection of these two galaxies in this image therefore makes their detection in the F336W image tentative at best. Object {\it GC1} has been already discussed by \citet{Vanzella2012}, who argue that the signal, believed to be LyC emission, is in fact emission from a $z \sim 1.6$ interloping galaxy. Finally, in Appendix \ref{emitter} we present a longer discussion regarding the galaxy {GC2}. We conclude that this galaxy is contaminated. Among the galaxies in our sample we therefore cannot claim a significant detection of Lyman continuum.

\subsection{Intergalactic medium}
\label{igm}

\begin{figure}
\centering
\includegraphics[scale=0.45]{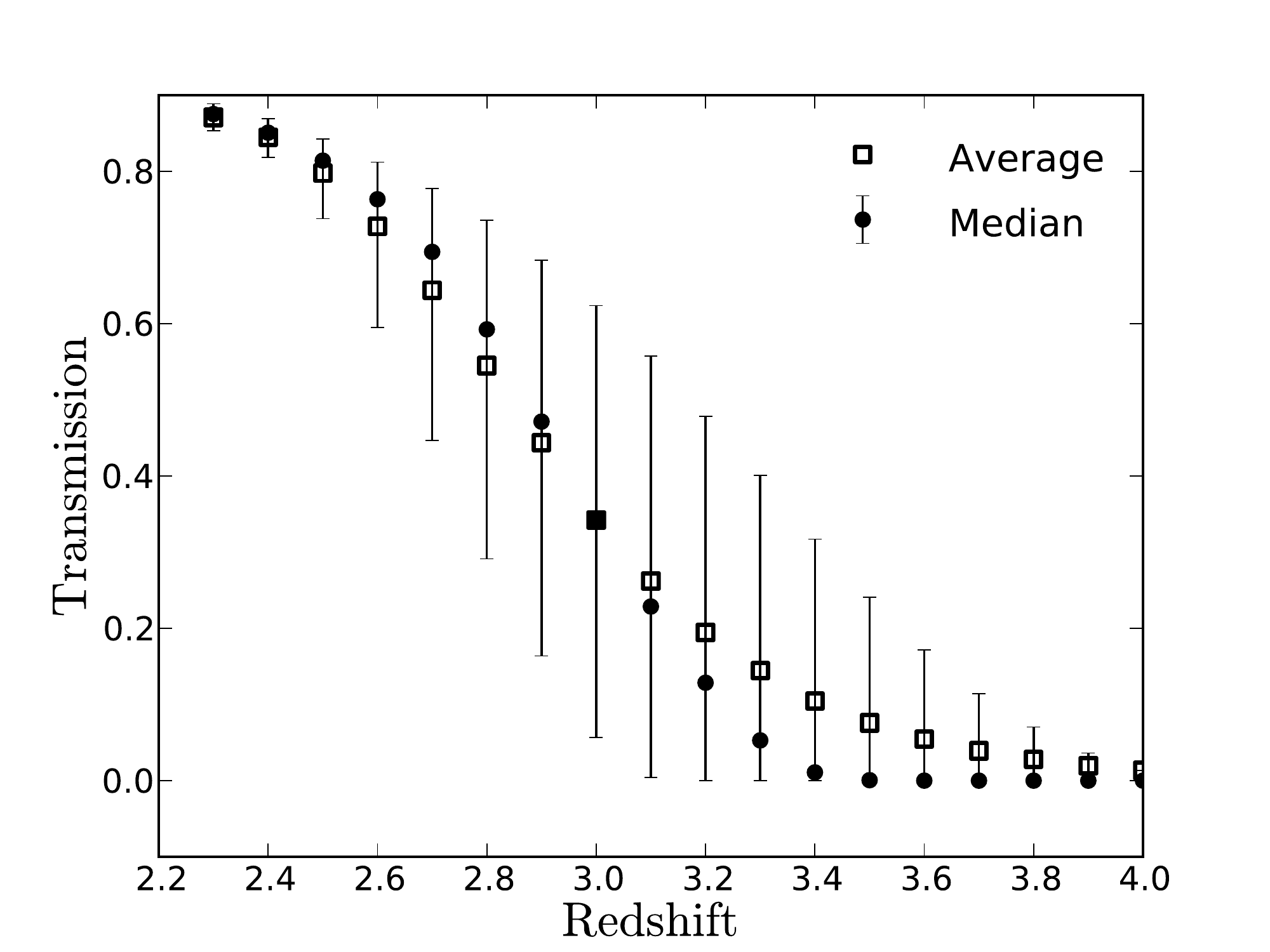}
\caption{Median (full circles) and average (empty squares) IGM transmission convolved with the {\it HST} WFC3-UVIS F336W filter as a function of redshift. Values were obtained by averaging 10000 lines of sight. Error bars indicate the 68$\%$ confidence level.}
\label{fig6}
\end{figure}

With increasing redshift the absorption due to IGM becomes a limiting factor in our ability to detect Lyman continuum emission. This is illustrated in Fig. \ref{fig0} where we show how the average transmission decreases for a galaxy at $z = 3.5$ with respect to the one at $z = 3$. To measure the real Lyman continuum that is being emitted by a galaxy, one has to apply proper correction to the measured fluxes (or, in the case of a non detection, to its upper limits). The absorption of the Lyman continuum is caused mainly by systems with relatively high \HI\, column densities $N_{HI} > 10^{-17}$ cm$^{-2}$ \citep{Inoue2008} and therefore the absorption is characteristically stochastic. Its effects have to be taken into account in a statistical manner by simulating a large number of lines of sight, based on the observed properties of the IGM absorbers. We calculate the transmission of the IGM by performing Monte Carlo (MC) simulation, following the prescription of \citet{Inoue2008}, \citet{Inoue2014} and \citet{Vanzella2015}. The empirical distribution functions of the redshift-dependent properties of intergalactic absorbers (e.g. \HI\, column density, number density, Doppler parameter) are used to generate a large number of absorbers along 10000 lines of sight at the given redshift. In Fig. \ref{fig6} we show the median and average (from now on $<T>$) transmission along 10000 simulated lines of sight in the $z = 2.3 - 4$ range, where the transmission is convolved with the shape of the F336W filter. The convolved average transmission drops below $<T> = 0.2, 0.1$ and 0.05 at $z = 3.2, 3.4$ and 3.6.

\subsection{Expected number of detections}
\label{measure}

Relative escape fraction $f_{\rm esc,rel}$, i.e. the fraction of escaping LyC photons relative to the fraction of escaping non-ionizing UV photons, can be written as \citep{Steidel2001,Siana2007}:
\begin{equation}
\label{eq1}
f_{\rm esc,rel}=\frac{\left(L_{\rm UV}/L_{\rm LyC}\right)^{\rm int}}{\left(F_{\rm UV}/F_{\rm LyC}\right)^{\rm obs}}\exp{\left(\tau_{\rm LyC}\right)},
\end{equation} 
where $\left(L_{\rm UV}/L_{\rm LyC}\right)^{\rm int}$ is the intrinsic luminosity density ratio, $\left(F_{\rm UV}/F_{\rm LyC}\right)^{\rm obs}$ is the observed flux density ratio and $\exp{\left(\tau_{\rm LyC}\right)}$ represents the line-of-sight opacity of the IGM for the Lyman continuum photons. In this work we do not work with the absolute escape fraction, i.e. relative escape fraction corrected for dust extinction $A_{\rm UV}$, $f_{\rm esc}=f_{\rm esc,rel}10^{-0.4A_{\rm UV}}$, due to the uncertainties involved in the measurements of the extinction. Note that the relative escape fraction is the quantity used to calculate the contribution from the observed UV LFs of high-z galaxies to the ionizing background.

In the previous section we have shown that we do not find a clear case with Lyman continuum emission among the galaxies in our sample. Indeed, from Eq. \ref{eq1} we see that, from purely observational perspective, both faintness of our sample and the high redshift range -- and therefore low IGM transmission -- work against us. Following \citet{Vanzella2010} we first simulate how many detections are expected (on average) for different values of assumed $f_{\rm esc,rel}$, given the intrinsic properties of our {\it clean} sample. The simulation is performed in the following way. We take the {\it clean} sample of 145 galaxies with known redshifts and $F_{\rm UV}$. For each galaxy in the sample we simulate 10000 different lines of sight with different IGM properties (see Section \ref{igm}), i.e. we simulate 10000 galaxy samples. For each galaxy in each sample we calculate the expected $F_{\rm LyC}$ by inverting Eq. \ref{eq1}. An error is assigned to the $F_{\rm LyC}$ by considering the dependency between the measurement error as a function of brightness, which has been obtained from the analytical fit to the measurements reported in the \citet{Rafelski2015} catalogue. If the expected flux is greater than the RMS, measured on the F336W image, the galaxy is counted as detected. The simulation is performed as a function of increasing $f_{\rm esc,rel}$. For the purpose of the simulation we made the following assumptions about the values of each quantity in Eq. \ref{eq1}:
\begin{itemize}
\item $\left( F_{\rm UV} \right)^{\rm obs}$ is derived from the observed $i_{775}$ magnitude of each source. The matching ACS/F775W filter corresponds to the rest-frame $\lambda_{\rm eff} \sim 1940, 1725$ and 1551 $\mathrm{\AA}$ at $z = 3, 3.5$ and 4, respectively.
\item $\left(L_{\rm UV}/L_{\rm LyC}\right)^{\rm int}$. Due to a lack of strong observational constraints the intrinsic luminosity density ratio is usually estimated from spectral synthesis models \citep[e.g.][]{Bruzual2003}. The values of the ratio in the literature are typically found in the range of $\left(L_{\rm UV}/L_{\rm LyC}\right)^{\rm int}\sim 2 - 9$ \citep[e.g.][]{Inoue2005,Siana2007,Siana2010,Nestor2013}, where the spread reflects different assumptions of the stellar population age, metallicity, star formation history and IMF. In general, younger stellar populations and lower metallicity will result in lower luminosity ratio. \citet{Izotov2016b} fitted synthetic models to the observed SEDs of the five low-$z$ galaxies with detected Lyman continuum and obtained $\left(L_{\rm UV}/L_{\rm LyC}\right)^{\rm int}\sim 1 - 1.3$. For the purpose of the simulation we assume the luminosity ratio to be distributed according to a gaussian distribution with the mean of {\it (a)} $\left<L_{\rm UV}/L_{\rm LyC}\right>^{\rm int}=7$, corresponding to the typical value for star-forming galaxies at our redshift range, and {\it (b)} $\left<L_{\rm UV}/L_{\rm LyC}\right>^{\rm int}=3$, corresponding to more extreme galaxies with young stellar populations. The standard deviation of the distribution is assumed to be 50$\%$ of the mean value. Our sample spans in a wide redshift range. Not only are the properties of the star-forming galaxy population expected to evolve in this range \citep[e.g.][]{Steidel2014,Barros2014}, but also the F336W filter probes bluer intrinsic spectral region with redshift, having an impact on the luminosity ratio that we should input into the simulation. Given the uncertainties in the estimated $\left<L_{\rm UV}/L_{\rm LyC}\right>^{\rm int}$, we do not consider these redshift-dependent effects.
\item IGM transmission in different lines of sight as a function of redshift is simulated as described in Section \ref{igm}. 
\item $f_{\rm esc,rel}$. It is reasonable to assume that the relative escape fraction is not the same in all galaxies but instead occupies a range of values. Due to the lack of actual measurements we don't know the shape of the actual distribution. Following \citet{Vanzella2010}, we run the simulation for three different distributions of $f_{\rm esc,rel}$: {\it (a)} constant value, {\it (b)} gaussian distribution, and {\it (c)} exponential distribution. Mean value for each distribution is varied between 0 and 1 in steps of 0.01. Note that from Eq. \ref{eq1} it follows that $f_{\rm esc,rel} < 1$ \citep{Vanzella2012}. A possible evolution of $f_{\rm esc,rel}$ with redshift and/or luminosity is not taken into account.
\end{itemize}

\begin{figure}
\centering
\includegraphics[scale=0.55]{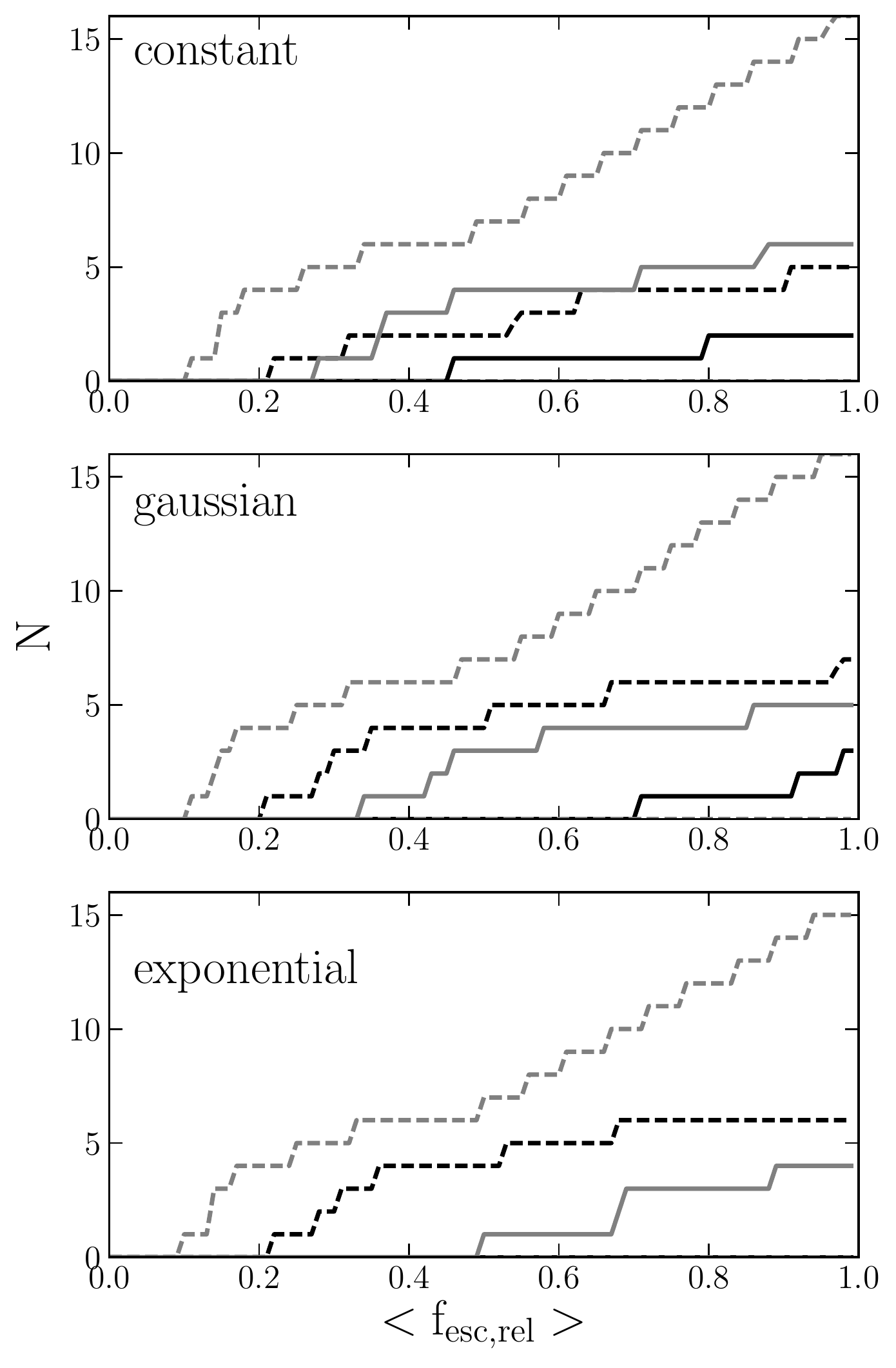}
\caption{Predicted number of Lyman leakers as a function of the assumed (average) relative escape fraction, given the properties of the {\it clean} sample of 145 sources. Three different escape fraction distributions have been considered in the MC simulation: constant, gaussian and exponential. Intrinsic luminosity density ratio $\left(L_{\rm UV}/L_{\rm LyC}\right)^{\rm int}$ has been assumed to have gaussian distribution with the mean value of 3 (grey lines) and 7 (black lines). Full lines represent the median value at each $<f_{\rm esc,rel}>$ step, and dashed line is the 1$\sigma$ equivalent upper limit.}
\label{fig7}
\end{figure}

The simulation is run separately for three different $f_{\rm esc,rel}$ distributions. Median and the 68$\%$ confidence level of the resulting relative escape fraction distribution at each $\left<f_{\rm esc,rel}\right>$ step is shown in Fig. \ref{fig7}. The results are shown only for the case where the Lyman continuum is detected at 2$\sigma$ significance. The results, though dependent on the simulation parameters, show that relatively high values of $f_{\rm esc,rel}$ at $3 < z < 4$ would be necessary in order to detect Lyman continuum in our sample of faint galaxies. 

\begin{table*}
\centering
\small
\renewcommand{\arraystretch}{1.5}
\begin{tabular}{lccccr}
\hline
\hline
\vspace{-0.2cm}
\multirow{2}{*}{$i_{\rm 775,cut}$} & \multicolumn{5}{c}{Redshift range ($<$T$>$)}\\
& 3.0 - 3.2 (0.26)& 3.2 - 3.4 (0.15)& 3.4 - 3.6 (0.08)& 3.0 - 3.6 (0.16)& 3.0 - 4.0 (0.11)\\
\hline

 24.0-26.0 & 0.07 (6)  & -9.0 (0)  & 0.58 (4)  & 0.11 (10) & 0.16 (13) \\
 26.0-27.5 & 0.34 (9)  & 0.53 (7)  & 1.29 (7)  & 0.31 (24) & 0.37 (32) \\
 27.5-28.5 & 0.73 (14) & 2.39 (5)  & 2.46 (13) & 0.78 (32) & 0.93 (44) \\
 24.0-27.0 & 0.1 (10)  & 0.52 (6)  & 0.59 (7)  & 0.15 (24) & 0.19 (33) \\
 27.0-28.5 & 0.51 (19) & 1.92 (6)  & 1.84 (17) & 0.57 (42) & 0.7 (56)  \\
 24.0-28.5 & 0.21 (29) & 0.71 (12) & 0.98 (24) & 0.25 (66) & 0.31 (89) \\
\hline
\end{tabular}

\bigskip

\begin{tabular}{lccccr}
\hline
\hline
\vspace{-0.2cm}
\multirow{2}{*}{$i_{\rm 775,cut}$} & \multicolumn{5}{c}{Redshift range ($<$T$>$)}\\
& 3.0 - 3.2 (0.26)& 3.2 - 3.4 (0.15)& 3.4 - 3.6 (0.08)& 3.0 - 3.6 (0.16)& 3.0 - 4.0 (0.11)\\
\hline

 24.0-26.0 & 0.21 (6)  & -9.0 (0)  & 1.74 (4)  & 0.34 (10) & 0.49 (13) \\
 26.0-27.5 & 1.02 (9)  & 1.59 (7)  & 3.88 (7)  & 0.92 (24) & 1.1 (32)  \\
 27.5-28.5 & 2.19 (14) & 7.18 (5)  & 7.39 (13) & 2.33 (32) & 2.78 (44) \\
 24.0-27.0 & 0.3 (10)  & 1.56 (6)  & 1.78 (7)  & 0.44 (24) & 0.58 (33) \\
 27.0-28.5 & 1.53 (19) & 5.77 (6)  & 5.52 (17) & 1.71 (42) & 2.11 (56) \\
 24.0-28.5 & 0.63 (29) & 2.12 (12) & 2.93 (24) & 0.75 (66) & 0.94 (89) \\
\hline
\end{tabular}
\caption{1-sigma (top) and 3-sigma (bottom) upper limits of relative escape fractions $f_{\rm esc,rel}$ for different redshift and luminosity bins. Average transmission in each redshift bin is reported in parentheses next to each redshift interval. Number of frames $N$ used in the stacking of each case is reported in parentheses in the table. Cases with $N<4$ are not considered in the analysis.}
\label{tab1}
\end{table*}

\subsection{Stacking analysis and $f_{\rm esc,rel}$ upper limits}
\label{stack}

Finally, we stack the F336W images of the galaxies in our sample to check whether we can detect a signal on a stacked image as well as to provide useful upper limits to the $f_{\rm esc,rel}$. As already discussed, the depth of the constraint on the escape fraction depends on the brightness of the galaxies and their redshift (i.e. IGM transmission). For this reason stacking the full sample is not very useful. We only consider the bright part of our sample ($i_{775} < 28.5$). Furthermore, galaxies are grouped into brightness and narrow redshift intervals, to illustrate how the $f_{\rm esc,rel}$ limit depends on the two quantities. 

\begin{figure}
\centering
\includegraphics[scale=0.55]{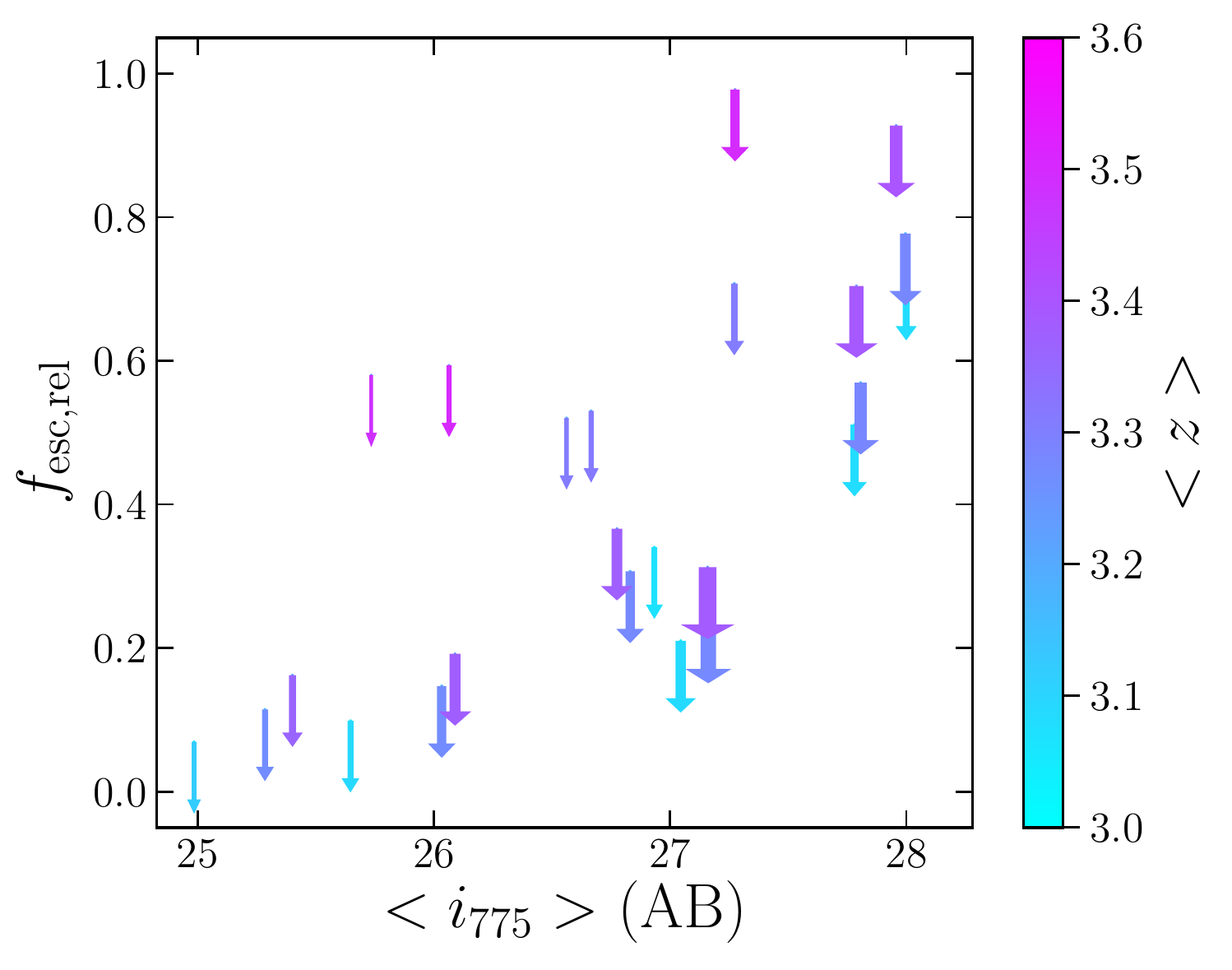}
\caption{Escape fraction 1$\sigma$ upper limits as reported in Table \ref{tab1} as a function of the apparent magnitude $i_{775}$. Colours indicate the average redshift of the galaxies in the subsamples used to derive each upper limits. Arrow widths are scaled according to the number of galaxies in each subsample. }
\label{fig8}
\end{figure}

First we apply a simple average stacking of the subsamples. The expected RMS of the stacked image is the 1$\sigma$ upper limit of 29.5 mag of the original image divided by $\sqrt{N}$, where $N$ is the number of images used in the stacking. Flux at the centre of each stacked image is measured within 0\farcs2 aperture and compared to the excepted RMS of the stacked image. We do not detect any signal with a significance greater than 1$\sigma$ in any of the stacked images. The upper limit of the $f_{\rm esc,rel}$ is estimated using Eq. \ref{eq1}, where we assume $\left(L_{\rm UV}/L_{\rm LyC}\right)^{\rm int} = 3$, $(F_{\rm UV})^{obs}$ is approximated with $i_{775}$, and the transmission is taken to be the average transmission in the corresponding redshift interval (see Figure \ref{fig6}). To further check our results we also apply different stacking techniques to the images: median stacking, weighted average stacking and weighted median stacking. The weights are not applied on the pixel-to-pixel basis, but rather image-wise: background noise in a region around each galaxy position is measured and its RMS is used as a weight for that particular galaxy. No significant flux is detected on any of the stacked images, regardless of the method used for stacking. Therefore we only consider simple average stacking in the following. The resulting $f_{\rm esc,rel}$  $1\sigma$ and $3\sigma$ upper limits are given in Table \ref{tab1}. The information in the table is graphically summarized in Fig. \ref{fig8}. As expected, escape fractions are less constrained when moving towards fainter galaxies: 1$\sigma$ limits increase from $\sim 0.07$ to 0.5 at $i_{775} \sim 25$ and 28 mag, respectively. Apparent is also the effect of the IGM transmission as, at a given brightness, escape fraction of the sample at lower redshifts are better constrained. As shown in the table, we do not profit anything by extending the sample to $z > 3.6$. 

It is interesting to note that the number of galaxies in the most constrained bins at faint luminosities is $\gtrsim 50$. Numerical simulations have shown that LyC radiation mostly escapes through narrow unobscured channels of star forming regions \citep[e.g.][]{Cen2015} and therefore the ability to detect LyC strongly depends on the viewing angle. The large number of galaxies we stack at faint luminosities is therefore much more representative as it would be if we had only a small number of galaxies.

\section{Discussion}
\label{discuss}

\subsection{{\it K}-band excess and Lyman continuum emission}
\label{excess}

\citet{Jaskot2013} and \citet{Nakajima2014} pointed out that there could be a connection between high \OIII/\OII\ and high $f_{\rm esc}$. In their scenario, high escape fractions are expected to arise in the case of density-bounded \HII\ regions with low \HI\ column densities. In the density-bounded scenario, the outer nebular regions in which the \OII\ is produced do not extend as far as in the case of ionization-bounded regions. At the same time the central nebular region, in which \OIII\ is produced, is expected to be of similar sizes in both cases. In such a situation one therefore expects higher \OIII/\OII\ ratios being connected to higher $f_{\rm esc}$. 

\begin{figure}
\centering
\includegraphics[scale=0.48]{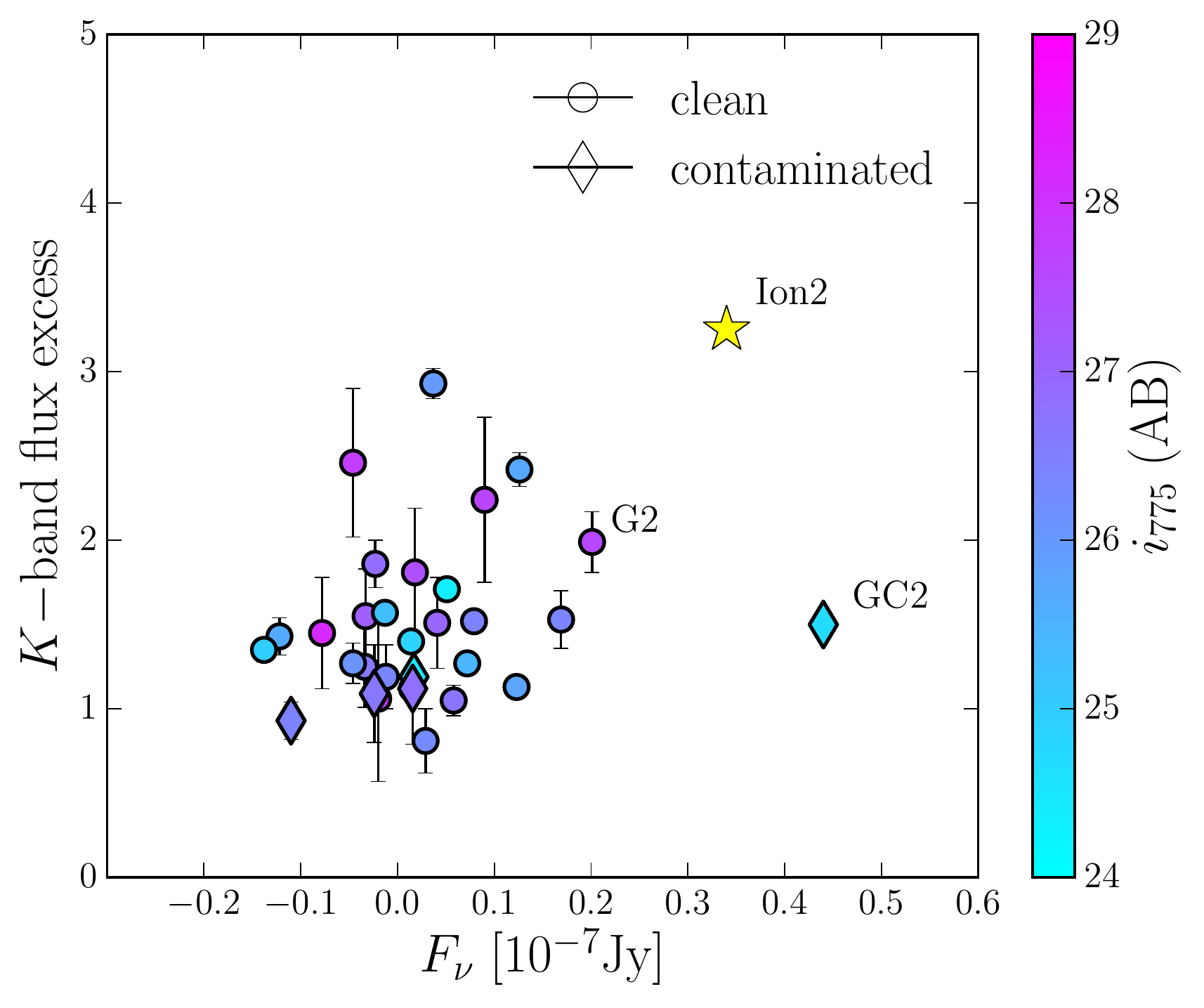}
\caption{Flux excess in the the {\it K}-band with respect to the fluxes in the adjacent photometric bands as a function of UVIS/F336W flux for 26 (5) galaxies in the {\it clean} ({\it contaminated}) sample. Included is the Lyman leaking {\it Ion2} galaxy \citep{Barros2016}.}
\label{fig10}
\end{figure}

The number of detected Lyman continuum leakers is still too small to allow a firm test of this relation. Still, galaxies with detected LyC show very strong \OIII\ lines both at low \citep{Izotov2016a,Izotov2016b} and high redshifts \citep{Vanzella2016}, corroborating this view (see also \citealt{Nakajima2016}). While we cannot claim a true detection of Lyman continuum in any of our galaxies, we nevertheless check the occurrence of objects with strong emission lines. From our {\it clean} and {\it contaminated} samples we look at the galaxies lying in the $3.0 < z < 3.6$ redshift range, for which the \OIII $\lambda4959,5007$ emission lines fall within the {\it K}-band. Then we measure the excess of the {\it K}-band flux with respect to the average flux of the adjacent photometric bands. At these redhifts the Balmer break falls within the {\it H}-band, therefore the ratio between the {\it H} and {\it K}-band alone traces the strength of the Balmer break instead of the strength between oxygen lines. For this reason, in addition to the measured {\it H}-band flux, we require the {Spitzer}/IRAC CH1 (i.e. 3.5 $\mu$m) observations. The average of the fluxes in these two bands should therefore reveal a strong {\it K}-band excess more accurately than if the comparison was made only with the {\it H}-band flux. Furthermore, the {\it Spitzer} measurements are required to be detected with at least 1$\sigma$ significance. For example, {\it G2} and {\it GC2} in Fig. \ref{fig5} qualify as good candidates. Limiting the analysis to $i_{775} < 28.5$, the ratio can be measured for 31 galaxies. We show the ratio as a function of the flux measured in the F336W in Figure \ref{fig10}. 

We find a large spread in the excess values. In all cases the excess is lower than the one found for {\it Ion2}. Our LyC candidate {\it GC2} has a marginal excess of $\sim 1.5$. For comparison, the five $z \sim 0.3$ LyC leakers found by \citet{Izotov2016b} would have an excess of $\sim 1.7 - 2.6$ if they were lying in our redshift range.

\subsection{Escape fraction as a function of luminosity}
\label{sim}

Our analysis of the faint sample of star-forming galaxies enabled us to extend the study of escape fraction down to $M_{1600} \sim -19$ mag at $z \sim 3 - 3.5$. The disadvantage of studying faint galaxies is that a large number of them is required in order to reach significant upper limits on the escape fraction. Consequently, strong observational constraints at the faint end are hard to reach. On the other hand, the luminosity dependence of the escape fraction has been suggested as a possible workaround to increase the contribution of star-forming galaxies to the ionizing background at high redshifts \citep{Kuhlen2012,Bouwens2012,Finkelstein2012,Fontanot2012,Fontanot2014,Robertson2013,Faisst2016}. In the following we consider the luminosity-dependent escape fraction using both our measurements and  a prediction by \citet{Duncan2015} with a goal to better understand the faint star-forming population that could provide necessary emission for ionization at high redshifts. 

\begin{figure}
\centering
\includegraphics[scale=0.55]{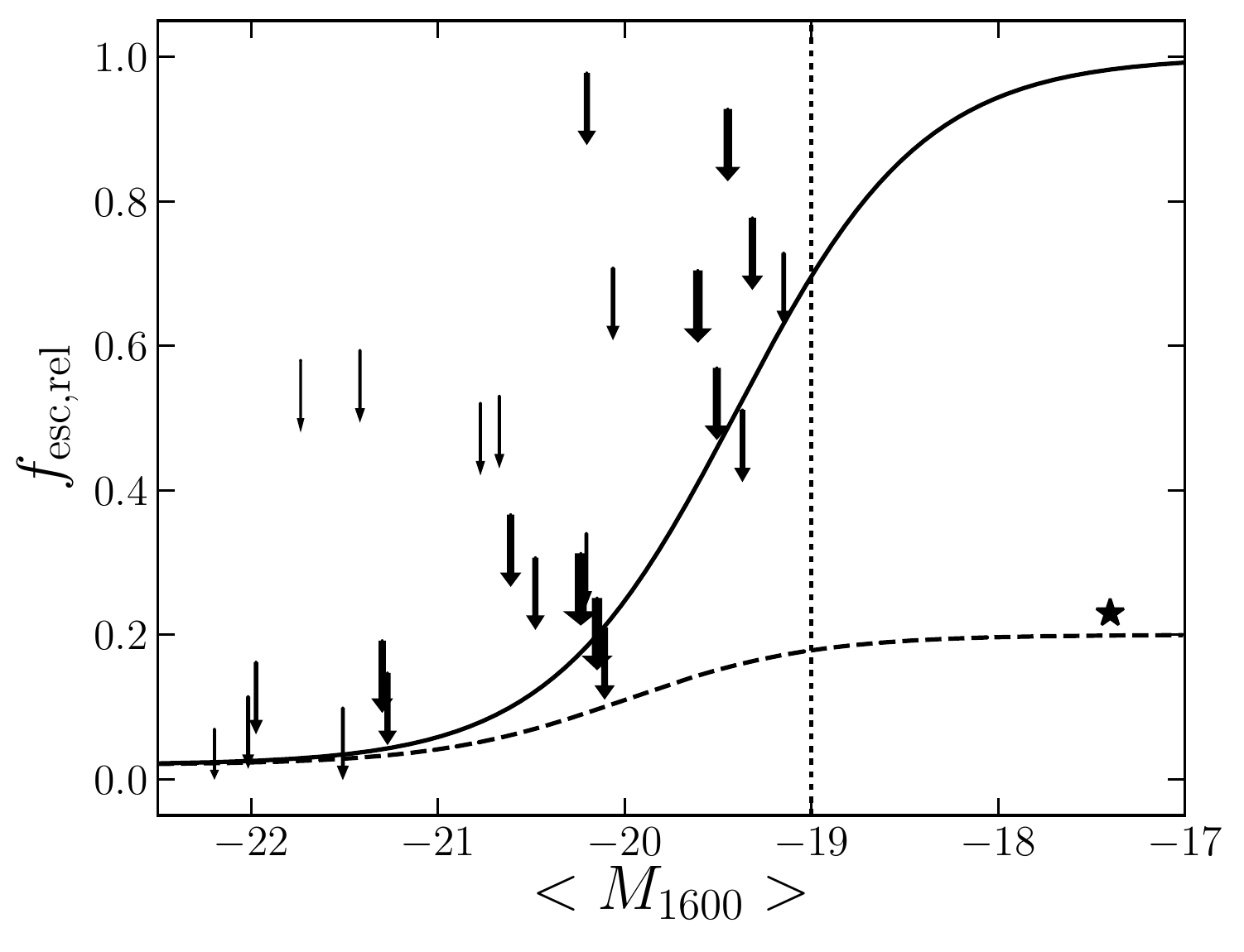}
\caption{Relative escape fraction upper limits as a function of absolute magnitude. Solid line represents an estimated upper limit envelope (Eq. \ref{eq2}). Dashed line shows the $f_{\rm esc}(M_{\rm UV})$ relation for which the modelling produces a background best matching with the data (see the text and Fig. \ref{fig13}, left panel, right column). Shown is also the escape fraction upper limit measured by \citet{Amorin2014} for a single faint galaxy (star).}
\label{fig12}
\end{figure}

Following the formalism of \citet{Fontanot2014} (see also \citealt{Cristiani2016}) we estimate the relative contribution of AGNs and star forming galaxies to the photon volume emissivity ($\dot{N}_{\rm ion}$) and photoionization rate ($\Gamma$) and compare the values to those determined by observations. In this paper we focus on the possible dependence of  $f_{\rm esc}$ on $M_{\rm UV}$ in a star forming population. The ionizing background associated with each population is modelled in the following way. The emission rate of hydrogen-ionizing photons per unit comoving volume is computed as 
\begin{equation}
\dot{N}_{\rm ion}(z) = \int_{\nu_{\rm H}}^{\nu_{\rm up}}\frac{\rho_{\nu}}{h\nu}d\nu,
\end{equation}
\begin{equation}
\rho_{\nu} = \int_{L_{\rm min}}^{\infty} f_{\rm esc}(L) \Phi(L,z)L_{\nu}(L)dL,
\end{equation}
where $\nu_{\rm H}$ corresponds to frequency at 912 $\mathrm{\AA}$, $\nu_{\rm up} = 4\nu_{\rm H}$ (e.g. we assume that photons with higher energies are mostly absorbed by \HeII; \citealt{Madau1999}), $\rho_{\nu}$ is monochromatic comoving luminosity density for sources brighter than $L_{\rm min}$, and $\Phi(L,z)$ is a redshift-dependent luminosity function. Note that for both populations we assume that the escape fraction does not vary with redshift. Photoionization rate is computed as \citep[e.g.][]{haardt2012}
\begin{equation}
\Gamma(z) = 4\pi \int_{\nu_{\rm H}}^{\nu_{\rm up}} \frac{J(\nu,z)}{h\nu}\sigma_{\rm \tiny \HI}(\nu) d\nu,
\end{equation}
where $\sigma_{\rm \tiny \HI}(\nu)$ is the absorption cross-section of neutral hydrogen. $J(\nu,z)$ is the background intensity:
\begin{equation}
J(\nu,z) = \frac{c}{4\pi} \int_{z}^{\infty} \varepsilon_{\nu_{1}}(z_{1})e^{-\tau_{eff}}\frac{\left(1+z\right)^3}{\left(1+z_{1}\right)^3} \left| \frac{dt}{dz}\right| dz_{1},
\end{equation} 
where the emissivity $\varepsilon_{\nu_{1}}$ is equivalent to $\rho_{\nu}$ in the comoving frame, $\nu_{1}=\nu\frac{1+z_{1}}{1+z}$ and $\tau_{\rm eff}=\tau_{\rm eff}(\nu,z,z_{1})$ is the effective optical depth for photons of frequency $\nu$ at $z$ that were emitted at $z_{1}$:
\begin{equation}
\tau_{\rm eff}(\nu,z,z_{1}) = \int_{z}^{z_{1}}dz_{2}\int_{0}^{\infty} dN_{\rm \tiny \HI} f(N_{\rm \tiny \HI},z_{2}) \left( 1 - e^{-\tau_{c}(\nu_{2})}\right).
\end{equation}
Here $\tau_{\rm c}$ represents the continuum optical depth through an individual absorber at frequency $\nu_{2}=\nu\frac{1+z_{2}}{1+z}$ and $f(N_{\rm \tiny \HI},z)$ is the distribution of absorbers. For the latter we adopt the distribution from \citet{Becker2013}. 

To calculate the AGN contribution to ionizing background we use the same approach as in \citet{Cristiani2016}. We approximate the AGN spectra with a power-law form, $f_{\nu} \sim \nu^{-\alpha}$, with $\alpha = 1.75$. For the AGN luminosity function we use the bolometric luminosity function prescription at $z<4$ given by \citet{Hopkins2007} and extrapolate it to higher redshifts. AGNs as faint as $L_{\rm min} = 0.1L_{\star}$ are taken into account: as shown by \citet{Cristiani2016}, objects fainter than that limit should provide a negligible contribution to the ionizing photon budget. Finally, we do not make the usual assumption of $f_{\rm esc}^{\rm AGN} = 1$ but rather adopt $f_{\rm esc}^{\rm AGN} = 0.75$, which is the average value measured for AGNs at $z \sim 3.6 - 4.0$ \citep{Cristiani2016}. We also assume that $f_{\rm esc}^{\rm AGN}$ does not depend on the luminosity in the considered luminosity range.

As for star forming galaxies, we use in Eq. (1) to (4) the evolution of Lyman Break Galaxy luminosity function (\citealt{Bouwens2015}, see also \citealt{Finkelstein2015}). Following \citet{Fontanot2014}, we estimate $L_{\rm min}(z)$ assuming that galaxy formation becomes inefficient in dark matter haloes smaller than the characteristic mass \citep{Okamoto2008}, i.e. that the contribution of fainter galaxies to the ionizing background is negligible. In detail, we use the characteristic mass estimate based on a blazar heating thermal history (see Fig 2. in \citealt{Fontanot2014}). Finally, we consider a redshift-dependent spectral emissivity as in \citet{haardt2012}.

\begin{figure*}
\centering
\begin{tabular}{lr}
\includegraphics[scale=0.43]{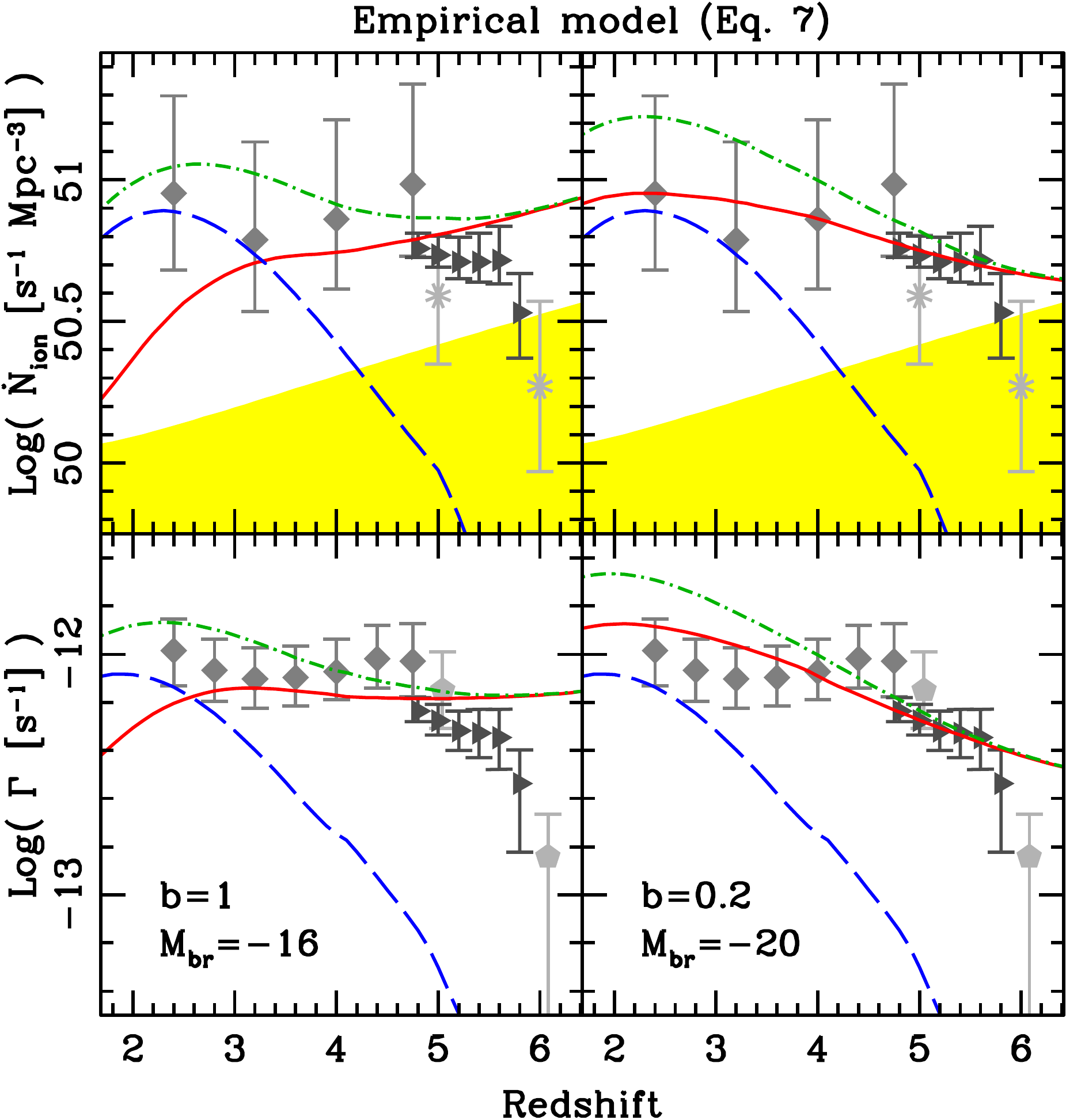}&
\includegraphics[scale=0.43]{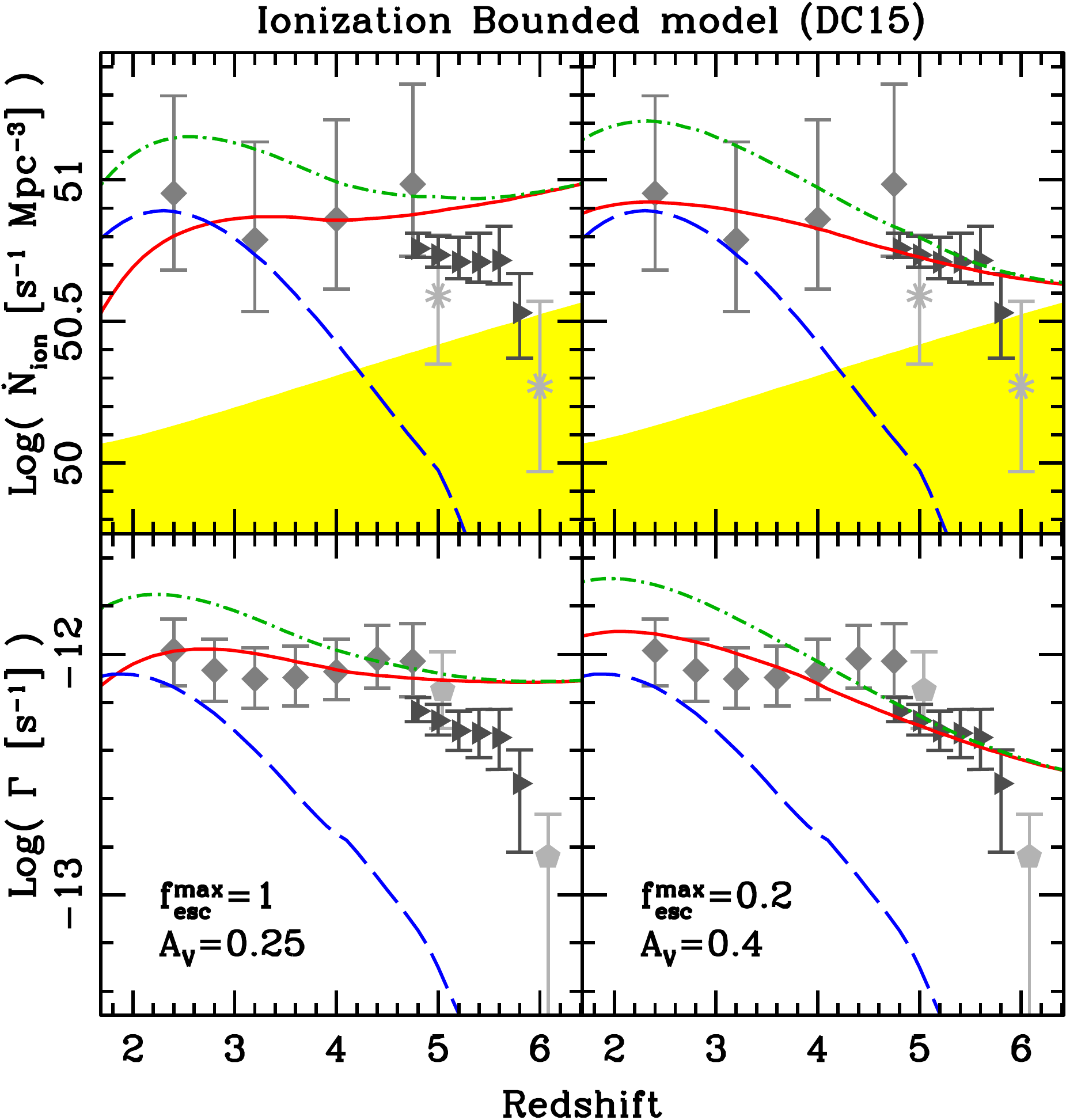}
\end{tabular}
\caption{Predicted photon volume emissivity (top) and photoionization rate (bottom) as a function of redshift. {\it Left:} $f_{\rm esc}(M_{\rm UV})$ empirical model with a transition in escape fraction (Eq. \ref{eq2}) from $f_{\rm esc} \sim 0$ to $f_{\rm esc} = b$ at break magnitude $M_{\rm br}$. {\it Right:} $f_{\rm esc}(M_{\rm UV})$ physical model arising in ionization-bounded nebula with holes (model A in Fig. \ref{fig14}; \citealt{Duncan2015}). The models are shown for two different highest allowed escape fractions of $f_{\rm esc,max} = 1$ and $f_{\rm esc,max} = 0.2$. Blue dashed curve shows the contribution from AGNs, red solid line from star-forming galaxies and green dotted-dashed line is the contribution from both. The upper edge of the yellow region denotes the minimum photon volume emissivity necessary for keeping the reionization. Measurements of photon volume emissivity are taken from \citet{Becker2013} (diamonds), \citet{Wyithe2011} (asterisks) and \citet{DAloisio2016} (triangles). Measurements of photoionization rate are taken from \citet{Becker2013} (diamonds), \citet{DAloisio2016} (triangles) and \citet{Calverley2011} (pentagons).}
\label{fig13}
\end{figure*}

In order to estimate the contribution of star forming galaxies to $\dot{N}_{\rm ion}$ and $\Gamma$ we consider different prescriptions for $f_{\rm esc}(M_{\rm UV})$\footnote{\citet{Sharma2016} uses the galaxies from the EAGLE cosmological hydrodynamical simulation \citep{Schaye2015} to estimate the redshift and luminosity dependence of the escape fraction by assuming that high average escape fractions are related to high star formation rate densities. Their prescription results in an escape fraction which increases with increasing UV luminosity. Their prediction at $z = 3$ (see Fig. 2 in \citealt{Sharma2016}) is at odds with the results presented in this paper as well as with the low upper limits found in previous works for the bright galaxies \citep[e.g.][]{Vanzella2010,Marchi2016}. In the following we therefore only consider models in which escape fraction increases with decreasing UV luminosity.}. As the first step we use our most constrained values (at each luminosity) to define an escape fraction upper limit envelope, which we represent with the following analytical function (see Fig. \ref{fig12}):
\begin{equation}
\label{eq2}
f_{\rm esc,rel}(M) = \frac{ae^{rM_{\rm br}} + be^{rM}}{e^{rM_{\rm br}} + e^{rM}},
\end{equation}
where $a = 0.02$ and $b = 1$ are high- and low-luminosity upper limits\footnote{The value of 0.02 is used as the faintest constraint measured at the bright end of luminosity function at $z \sim 1.3$ \citep[e.g.][]{Siana2007}. Nevertheless, as discussed later in the text, the results are largely insensitive to this value.}, $M_{\rm br} = -19.3$ mag is the transitional luminosity and $r = 2$ mag$^{-1}$ describes the smoothness of the transition. The parameters have been determined by matching the function to the data. Given that the actual shape of the $f_{\rm esc}(M_{\rm UV})$ is unknown, in the following we do not make any distinction between relative and absolute escape fraction in the modelling.

The results of our calculation using different models (introduced in the following text) are presented in Fig. \ref{fig13}, where they are compared to the measurements of $\dot{N}_{\rm ion}$ and $\Gamma$ from the literature. For each model we plot the contribution to the background from AGNs (dashed blue line), star forming galaxies (solid red line) and the contribution from both (dot-dashed green line). 

The functional form of $f_{\rm esc}(M_{\rm UV})$ assumed above is not physically motivated, i.e. it is based on our inability to make stringent measurements for the faint galaxy population. Nevertheless, we use it as a first crude approximation of the luminosity dependency. Taking this $f_{\rm esc}(M_{\rm UV})$ at face value, the calculated ionizing background turns out to be too high. By varying the transitional magnitude $M_{\rm br}$ we search for the best match between the calculated and observed $\dot{N}_{\rm ion}(z)$ and $\Gamma(z)$ and find that this is achieved with $M_{\rm br} \sim -16$ (see left column in the left panel in Fig. \ref{fig13}). Given the limiting integration luminosity $L_{\rm min}(z)$ \citep{Fontanot2014}, the background at $z \sim 4$ is mainly contributed by faint galaxies with $M_{\rm UV} \sim M_{\rm br}$ and is being dominated by progressively fainter population when moving towards higher redshifts. At $z \gtrsim 5$ we clearly start to overestimate the rate of emitted photons per unit volume. This is a combination of two effects: {\it (i)} the decrease in $L_{\rm min}$ with redshift which increases the contribution of faint galaxies with high escape fraction, and {\it (ii)} the progressively steeper galaxy luminosity functions at higher redshifts \citep{Bouwens2015} which increases the fraction of faint galaxies that have the highest escape fraction (in our model). 

While due to the lack of knowledge of the actual $f_{\rm esc}(M_{\rm UV})$ shape the hypothesis of a sharp transition in $f_{\rm esc}$ at some magnitude cannot be rejected, it is reasonable to assume that low-luminosity galaxies cannot have $f_{\rm esc} \sim 1$ on average. It is interesting to see what happens when the highest escape fraction allowed is lowered. \citet{Amorin2014} estimated an upper limit of $f_{\rm esc,rel}=23\%$ for a lensed, faint galaxy ($M_{\rm UV} = -17.4$ mag). This is by far the faintest galaxy for which the escape fraction has been constrained, but we caution that the number is very uncertain due to the uncertainties in the IGM transmission, and the result cannot be assumed to be representative of the whole galaxy population \citep{Cen2015}. Still, we set $f_{\rm esc,max} = 0.2$ based on that result and also to compare our results to other studies where the same value is often assumed. We find that the calculated ionizing photon production rate in this case is in better agreement with measurements at high redshifts (the right column of the left panel in Fig. \ref{fig13}). In this case the transitional magnitude for the model best matching the data is $\sim -20$ mag, which is significantly brighter than in the previous case. It is worth stressing that the strong drop in the observed UV background when approaching $z \sim 6$ is hard to reproduce within our current scheme.

Irrespective of the highest allowed escape fraction, the bright part of the galaxy population contributes a negligible amount of photons for reionization. In fact, the above results do not change appreciably if we change the high-luminosity upper limit parameter to $a = 0.05$ or $a = 0$. This means that even if observationally we could constrain the escape fraction of the bright population to very low upper limits (or low absolute values if in the future the stacking analysis of a large sample results in a significant signal), that alone would not help us to constrain the properties of the faint population that is the dominant contributor to the reionization.

\subsection{Comparison with a physically motivated model of the escape fraction}
\label{physical}

\begin{figure}
\centering
\includegraphics[scale=0.58]{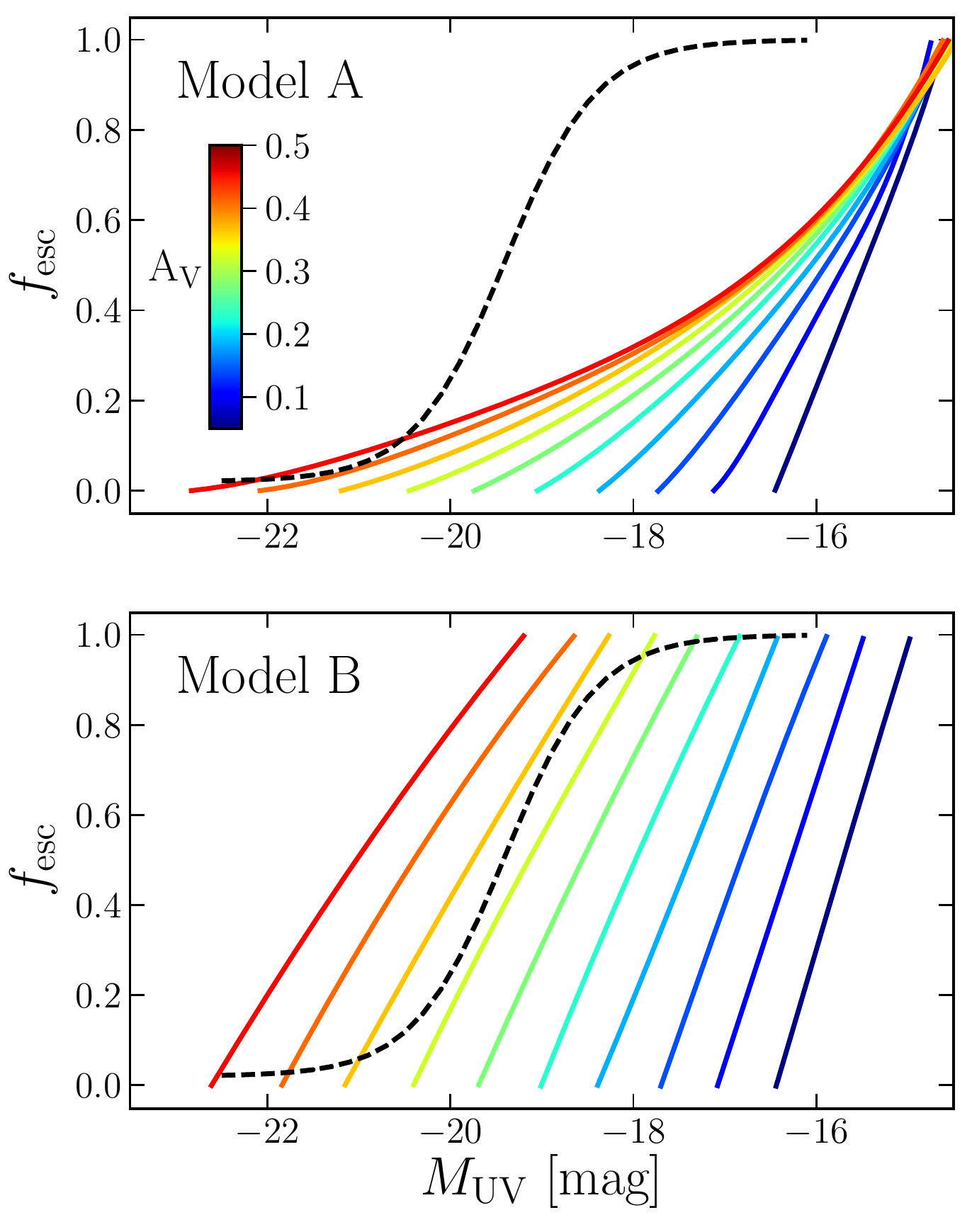}
\caption{Absolute escape fraction as a function of luminosity as predicted by \citet{Duncan2015} for the case of ionization-bounded nebula with holes (model A) and density bounded nebula (model B). Different curves correspond to different assumed values of dust extinction $A_{\rm V}$ of the dust screen enclosing a star-forming region. Black dashed curve is the upper-limit envelope of relative escape fraction constrained by our measurements.}
\label{fig14}
\end{figure}

Robust constraints at the faint end of $f_{\rm esc}(M_{\rm UV})$ are observationally hard to achieve and, as illustrated in the discussion so far, the study of the $f_{\rm esc}(M_{\rm UV})$ is limited to a qualitative analysis. Motivated to explore the issue further, we repeat the above analysis but this time using a physically motivated model for $f_{\rm esc}(M_{\rm UV})$. \citet[][hereafter DC15]{Duncan2015} looked into the properties of LyC escaping from two particular models of star-forming regions: {\it (A)} a ionization-bounded nebula with dust-free holes, through which the LyC can escape; and {\it (B)} a density-bounded nebula where LyC can escape due to the incomplete Str\"omgren sphere (see also \citealt{Zackrisson2013}). Under certain assumptions on stellar population age, star-formation history, metallicity and dust extinction (see DC15 for details), they used a stellar population synthesis analysis to get relations between "observed" UV slope $\beta$, absolute escape fraction and extinction $A_{\rm V}$ for each of the two models. We use the results reported in their Figs. 5 and 6 to extract the $f_{\rm esc}(\beta)$ relations for different $A_{\rm V}$ values. DC15 compiled observational relations between UV spectral index $\beta$ and $M_{\rm UV}$ at different redshifts. We use the relation at $z \sim 4$ to transform the $f_{\rm esc}(\beta)$ to $f_{\rm esc}(M_{\rm UV})$, as the relation at this redshift is the most reliable over a wide luminosity range. The resulting relations for the two models are presented in Fig. \ref{fig14} and compared with the empirical model. We then repeat the analysis as in the previous section using these relations as input. We note that there are a few caveats when using the extracted models. The analysis of DC15 specifically focused on $z > 6$ galaxies: this reflects in the choice of the assumed parameters in their spectral synthesis analysis. In addition, we do not take into account the variation in the $\beta (M_{\rm UV})$ relation with redshift (DC15; \citealt{Bouwens2014,Rogers2014}). Nevertheless, as long as we carry these caveats in mind, the extracted $f_{\rm esc}(M_{\rm UV})$ models should provide some valuable information, given the number of uncertainties and assumptions involved in the modelling of the ionizing background on the one hand and in the stellar synthesis analysis of DC15 on the other.

We redo the calculation of the ionizing background for the two models of DC15. Each model is limited to $f_{\rm esc,max}=1$ at the faint end. Both models in principle give very similar results to the ones obtained by our starting model, provided that the $A_{\rm V}$ is sufficiently low. Model B $f_{\rm esc}(M_{\rm UV})$ relations have a rather sharp transition from $f_{\rm esc} = 0 - 1$, and the calculated ionizing background is qualitatively similar to the one we obtained from our initial modelling using Eq. \ref{eq2}. Therefore we only show the calculated background using the model A relations (right panel in Fig. \ref{fig13}). Again we obtain a similar overestimation of the background at high redshifts, which we attribute to the very high values of escape fraction at low luminosities. Fixing the highest allowed escape fraction to $f_{\rm esc,max} = 0.2$ results in a flat evolution all the way to the highest redshifts (see second column of the right panel in Fig. \ref{fig13}).

It is interesting to see that our measured upper limits can put some constraints to the parameter space inspected by DC15. Dust effects are understandably much stronger in the density-bounded model (model B). High dust extinction favours a relevant contribution of bright galaxies to the ionizing photons, which is at odds with our observational limits. On the other hand, based on the high redshift measurements of $\dot{N}_{\rm ion}$ and $\Gamma$ it is clear that galaxies cannot have very high escape fraction on average, as that leads to an overprediction of the background. Our "sharp" transition model with $f_{\rm esc,max} = 0.2$ shows that the transition is expected at $M_{\rm UV} \sim -20$ mag, but could also be in the somewhat less luminous regime. Increasing the number of observed galaxies with $M_{\rm UV} \in (-20,-18)$ mag is therefore critical to assess the level of importance of faint galaxies in reionization, irrespective of the actual $f_{\rm esc}(M_{\rm UV})$ relation. Finally we emphasize that we did not include any evolution of the escape fraction with redshift in the modelling. As shown in the past \citep[e.g.][]{Kuhlen2012,Fontanot2012}, an escape fraction independent of luminosity, but increasing with redshift, could likewise provide enough photons for ionization. Unfortunately, the information of a possible $f_{\rm esc}(z)$ evolution can only be inferred from secondary means (e.g. change in other galaxy properties as the emission line strength) or simulations \citep[e.g.][]{Sharma2016}. 

\section{Conclusions}

As the average escape fraction of hydrogen ionizing photons in star forming galaxies at the bright end of luminosity function at $z \sim 3$ is found to be low and in contrast with expectations based on the observed properties of reionizing background, it has been often suggested that the escape fraction may increase with decreasing luminosity of galaxies. In this paper we have studied this hypothesis by analysing the escape fraction of relatively faint galaxies. We combined the deep ultraviolet observations of the Ultra Deep Field with deep spectroscopic MUSE observations of the same field in order to compile a sample of $0.02 \lesssim L/L_{\rm z=3}^{*} \lesssim 10$ galaxies at $z > 3$ with photometric observations of their $\lambda < 912 \mathrm{\AA}$ part of the spectrum. None of the $165$ galaxies in our sample shows a significant Lyman continuum emission (Section \ref{results} and Appendix \ref{emitter}).

Limiting our analysis to the brighter half of the sample ($L \gtrsim 0.1 L_{\rm z=3}^{*}$) we use stacking analysis to provide deep upper limits on the relative escape fraction down to $M_{\rm UV} \sim -19$ mag (Section \ref{stack}). Stacking is performed for several subsamples of galaxies lying at different redshift and brightness intervals. We do not detect any significant LyC signal in any of the stacked images but provide $1\sigma$ upper limits of $f_{esc,rel} < 0.07, 0.2$ and 0.6 at $L \sim L_{\rm z=3}^{*}, 0.5L_{\rm z=3}^{*}$ and $0.1L_{\rm z=3}^{*}$, respectively.

Our measurements allow us to constrain the $f_{\rm esc}(M_{\rm UV})$ relation for star forming galaxies. Under the assumption that AGNs and star forming galaxies are the only sources of reionizing background, and using various prescriptions for $f_{\rm esc}(M_{\rm UV})$, we estimate the relative contribution of AGNs and star forming galaxies to the photon volume emissivity and photoionization rate and compare the values to those determined by observations (Section \ref{sim}). We find that the bright part of the galaxy luminosity function (e.g. $M_{\rm UV} < -20$ mag), given the low upper limit constraints, provides a negligible amount of reionizing photons, irrespective of the shape of the $f_{\rm esc}(M_{\rm UV})$ relation in the faint part. We also show that measurements like ours can provide means to better understand the general properties of star forming regions (Section \ref{physical}). For example, comparing our upper limits with predictions from physical models \citep{Duncan2015}, we show that density-bounded nebulae with $A_{\rm V} >0.3$ are disfavoured. Deeper escape fraction constraints down to $M_{\rm UV} \sim -18$ mag are therefore required to better understand both the $f_{\rm esc}(M_{\rm UV})$ relation as well as star forming regions at high redshifts.

For a more quantitative analysis we need more stringent constraints on the fainter part of the luminosity function, which could be achieved by repeating our analysis using a larger sample of faint galaxies. With deeper MUSE observations on this field, to become available in the future, it will be possible to greatly expand the number of spectroscopically confirmed faint Lya emitters.
Moreover, expanding the UVUDF field to cover larger area coinciding with wider, but shallower, MUSE coverage of the UDF field (i.e. GTO program ID 095.A-0240(A), PI: L. Wisotzki) will be essential to better constrain the escape fraction dependency on the luminosity. Alternatively, the low-luminosity regime can be more effectively probed by taking advantage of the gravitational lensing amplification effect by massive galaxy clusters with multi-band HST coverage, such as the programs CLASH, Hubble Frontier Fields and RELICS \footnote{https://archive.stsci.edu/prepds/relics/}. In any case, understanding the role of star forming galaxies in the process of reionization at high redshifts will require a joint effort of direct searches for Lyman continuum leakers and improving our detailed knowledge of star forming regions and their evolution with redshift. 

\section*{Acknowledgements}
We thank the anonymous referee for a helpful report which improved this manuscript. We acknowledge financial contribution from the grant PRIN-MIUR 2012 201278X4FL 002 The Intergalactic Medium as a probe of the growth of cosmic structures. We acknowledge financial contribution from the grant PRIN-INAF 2014 1.05.01.94.02. 


\bibliographystyle{mnras}
\bibliography{ms_japelj} 

\appendix

\section{Galaxy {\it GC2}}
\label{emitter}

Galaxy {\it GC2} (GOODS ID: J033240.30-274752.6) represents an interesting case. It is significantly detected both in the F336W and F275W image, and with less significance in the F225W images \citep{Rafelski2015}. This is surprising, as one would expect the signal to drop towards blue due to both the intrinsic galaxy emission properties and, on average, decreasing IGM transmission (see Fig. \ref{fig0}). For example, at the redshift of this galaxy ($z = 3.45$), the convolved IGM transmission with the F275W filter results in an average transmission of $<T>\sim 0.045$, approximately half of the convolved transmission through the F336W filter. The SED reveals a slight {\it K}-band excess indicating strong $\OIII\lambda4959,5007$ emission lines. The MUSE spectrum reveals a double-peaked Ly$\alpha$ lines separated by $\sim 700$ km s$^{-1}$. \citet{Verhamme2015} used Ly$\alpha$ radiation transfer calculations in the H II regions and showed that a peak separation of $<300$ km s$^{-1}$ is favoured in the case of galaxies with escaping LyC radiation. The value for {\it GC2} is therefore much higher than predicted for the case of escaping LyC. We note that for a few galaxies among the confirmed LyC leakers the peak separations of a double-peaked Ly$\alpha$ line were actually significantly higher than the predicted limit of 300 km s$^{-1}$ \citep{Barros2016,Verhamme2016}, even though not as large as in our present case. The rest-frame Ly$\alpha$ equivalent width (EW) is estimated to be EW $\approx 40~\mathrm{\AA}$. 


\begin{figure}
\centering
\includegraphics[scale=0.18]{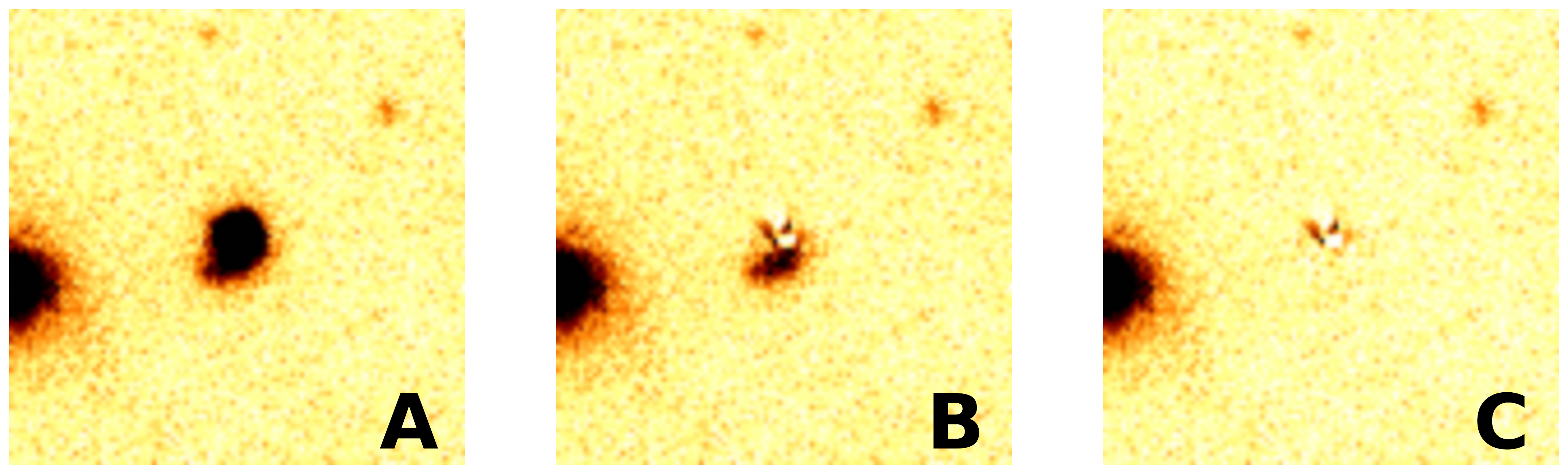}
\caption{{{\sc Galfit} modelling of the {\it GC2} galaxy in the F775W band. Shown are the original image (A), the image after subtracting the main galaxy component (B) and the image after subtracting both galaxy components (C).}}
\label{fig15}
\end{figure}

The optical image of the galaxy shows a structured morphology with a bright compact source and a faint tail. The UV emission comes exclusively from the fuzzy part of the object. This feature, together with detection in all three UV bands, may indicate that we are dealing with a low-redshift interloper. The MUSE spectrum is rather noisy and we cannot confidently identify any absorption lines that would correspond to an interloper. We also do not detect any  emission lines (other than Ly$\alpha$) in the spectrum. This indicates that there may be no interlopers at $z < 1.5$, though the non-detection could also be simply due to the faintness of the interloper. Indeed, unless the interlopers have very strong emission lines, their faint nature presents a great challenge for current telescopes \citep[e.g.][]{Vanzella2012}. Observations in the optical and near-infrared with the future generation of extremely large telescopes will be necessary to provide consensus for many ambiguous sources found in this and previous studies.

Assuming that the faint tail is indeed another galaxy, we determine its brightness in the rest-frame UV band by modelling the {\it GC2} source in the HST ACS/F775W image with two galaxy components using {\sc GALFIT} \citep{Peng2010} and following the same procedure described in \citet{Vanzella2015}. The two, bright and faint, components are reasonably well modelled (as illustrated by the residual image in Figure \ref{fig15}). The bright component is found to be compact and marginally resolved (with an effective radius $\lesssim$200 pc). The faint component, which is of the main interest here, is well reproduced with a two-dimensional nearly Gaussian shape with a total magnitude of $i_{\rm 775} =27.1 \pm 0.1$ mag. On the other hand, the brightness in the F336W image is measured to be $27.5 \pm 0.2$ mag. Using these two measurements and Equation \ref{eq1} it follows that $f_{\rm esc,rel}>1$ for practically any physically-plausible IGM LOS. As pointed out by \citet{Vanzella2012}, the relative escape fraction from a particular part of the galaxy cannot be larger than 1. A value of $f_{\rm esc,rel}>1$ is therefore a strong argument in favour of contamination. Furthermore, the fuzzy part is also detected in the F275W: repeating the analysis for this band would result in an even larger values of $f_{\rm esc,rel}$.

Based on the presented analysis we conclude that the {\it GC2} source is contaminated. This case points to the importance of the high-resolution imaging in such studies, as already shown by previous works \citep{Vanzella2010,Siana2015,Mostardi2015}: the source would be classified as a Lyman continuum emitter if only the seeing-limited VIMOS-U image were available to us.

\bsp	
\label{lastpage}
\end{document}